# Speech Emotion Recognition Using Deep Sparse Auto-Encoder Extreme Learning Machine with a New Weighting Scheme and Spectral/Spectro-Temporal Features Along with Classical Feature Selection and A New Quantum-Inspired Dimension Reduction Method

Fatemeh Daneshfar, Seyed Jahanshah Kabudian[1]

Department of Computer Engineering and Information Technology,

Razi University, Kermanshah, IRAN.

## ABSTRACT

In today's world, affective computing is very important in the relationship between man and machine. In this paper, a multi-stage system for speech emotion recognition (SER) based on speech signal is proposed, which uses new techniques in different stages of processing. The system consists of three stages: feature extraction, feature selection/dimension reduction, and finally feature classification. In the first stage, a complex set of long-term-statistics features is extracted from both the speech signal and the glottal-waveform signal using a combination of new and diverse features such as prosodic features, spectral features, and spectro-temporal features. One of the challenges of the SER systems is to distinguish correlated emotions. These features are good discriminators for speech emotions and increase the SER's ability to recognize similar and different emotions. The data augmentation technique is also used to increase the number of training samples. This feature vector with a large number of dimensions naturally has redundancy. In the second stage, using classical feature selection techniques as well as a new quantum-inspired technique to reduce the feature vector dimensionality (proposed by the authors), the number of feature vector dimensions is reduced. In the third stage, the optimized feature vector is classified by a weighted deep sparse extreme learning machine (ELM) classifier. The classifier performs classification in three steps: sparse random feature learning, orthogonal random projection using the singular value decomposition (SVD) technique, and discriminative classification in the last step using the generalized Tikhonov regularization technique. Also, many existing emotional datasets suffer from the problem of data imbalanced distribution, which in turn increases the classification error and decreases system performance. In this paper, a new weighting method has also been proposed to deal with class imbalance, which is more efficient than existing weighting methods. The proposed method is evaluated on three standard emotional databases EMODB, SAVEE, and IEMOCAP. According to our latest information, the system proposed in this paper is more accurate in recognizing emotions than the latest state-of-the-art methods.

## 1- Introduction

Recognition of emotions from speech signal or Speech Emotion Recognition (SER) is one of the research fields in affective computing. The purpose of the SER is to analyze human speech signal and to extract the emotional state of the person (fear, joy, sadness, etc). The SER systems usually have three stages: feature extraction, dimension reduction/feature selection, and classification. In the feature extraction stage, prosodic and spectral features are usually used. These features alone are not able to discriminate unstable transitions in the speech signal in different emotions. For this purpose, in this paper, spectro-temporal features such as Gabor filter bank (GBFB) and separate Gabor filter bank (SGBFB) features [1] have been used, which have more discriminative power. New spectral features such as constant-Q cepstral coefficient (CQCC) [2], Single frequency cepstral coefficient (SFCC) [3] and IIR-CQT Mel-frequency cepstral coefficient (ICMC) [4] that have not previously been used to identify

---

[1] S.J. Kabudian

Department of Computer Engineering and Information Technology, Razi University, Kermanshah, Iran.
e-mail: Kabudian@razi.ac.ir *Corresponding Author*

emotion, have also been used. The set of prosodic, spectral and spectro-temporal feature vectors becomes a very large vector and long-term statistics are extracted from it. One of the most important issues in pattern classification is feature selection or dimension reduction. Unfortunately, classification error is not directly considered as cost function in current feature selection/dimension reduction algorithms. In this paper, in addition to using the classical feature selection methods, a new meta-heuristic method developed by the authors [5,6] and inspired by quantum theory- called point-mass quantum particle swarm optimization (pQPSO)- is used to estimate the dimension reduction matrix. After dimension reduction, the feature vector must be classified. This paper uses a deep classifier. One of the problems with deep learning is that they take a long time to train. The hierarchical adaptive weighted ELM (H-AWELM) classifier is used here, which is very fast. This classifier uses both random unsupervised feature learning as well as random orthogonal projections and supervised learning. This classification strategy is very fast because it uses random mappings in the unsupervised layers and the fast-regularized weighted pseudo-inverse (generalized Tikhonov regularization) technique in the supervised layer. Since emotional databases are typically imbalanced, a new sample-wise weighting method is proposed which is more accurate than the current weighting methods.

Also, in general, different speech emotions may have similar acoustical features. The correlation coefficient index can be used to examine the degree of dependence and relationship between different emotions. Fig. 1 shows a correlation diagram of seven different emotions of the EMODB dataset. From this diagram, it can be concluded that *sad* has the highest similarity with *bore* (compared to other emotions), *happiness* has the highest similarity with *anger* and *fear* has the highest correlation with *disgust* among other emotions. The reason for this is, the similarity in energy, pitch, duration, and other prosodic features in these emotions. Also, the pairs (*fear*, *anger*) and (*disgust*, *happiness*) have a low correlation coefficient and are independent. Recognizing and differentiating these correlated emotions is one of the challenges of today's speech emotion recognition systems, which in this study has been done well.

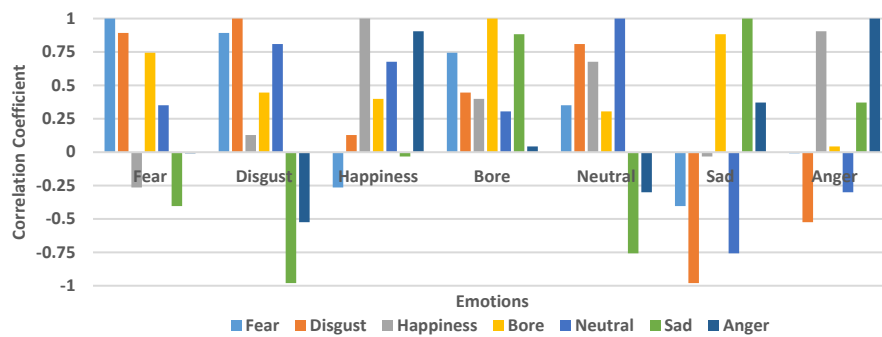

Fig. 1. Comparison of emotion correlation rates using correlation coefficient index

The importance of the results of this paper and its contributions are summarized as follows:
- In the feature extraction stage, spectro-temporal features such as GBFB and SGBFB are used, as well as new spectral features such as CQCC, SFCC, and ICMC, which have not yet been used in SER. These features help to better identify similar emotions with the same and different speakers by providing spectro-temporal contrast at all times. They will also improve the results of speech/non-speech classification by eliminating redundant information between feature components [1].
- In the feature dimension reduction stage, both classical methods and a new quantum-inspired meta-heuristic algorithm proposed by the authors are used to estimate the projection matrix.
- In the classification stage, the H-AWELM classifier is used along with a new weighting method. This new weighting method proposed by the authors leads to higher performance on three popular databases than the current weighting methods.
- According to our latest information, the proposed speech emotion recognition system has the best accuracy for the two well-known datasets EMODB and SAVEE compared to all other state-of-the-art systems.

The rest of the paper is organized as follows: Section 2 will review the research work done in the field of the SER. Section 3 describes the SER system, the algorithms and block details. The results of the experiments and performance evaluation of the proposed system, are presented in Sections 4 and 5 respectively, and finally we will conclude in the last section.

## 2- Literature Review

Numerous works have been recently published on emotional speech feature extraction. In [7], a new architecture has been proposed that extracts Mel-frequency cepstral coefficients, Mel spectrogram, and spectral contrast features from audio files and uses them as input to a deep convolutional neural network (CNN) to identify the emotions of the EMODB, IEMOCAP, and RAVDESS datasets. In this study, to improve the accuracy of classification, an incremental method has been used to modify the initial model. The final recognition rates obtained by this model were 86.1% and 64.3% on the EMODB and IEMOCAP datasets. In this study, correlated emotions (*happiness*, *anger*) are well differentiated (100%, 100%), while (*bore*, *sad*) and (*disgust*, *fear*) are not well distinguished (76.92%, 87.5%) and (90%, 66.67%) and classification error occurred.

The model presented by [8], extracts important discriminative salient features in parallel and learns long-term conceptual dependencies from speech, using a multi-learning strategy. In this study, deep learning (DL) is used to find the speech emotional features. This architecture consists of a Dilated CNN (DCNN) to learn the basic features of the input speech signal. It then uses one-dimensional residual blocks (RB) and a Bidirectional gated recurrent unit (BIGRU) to improve learned features. Also, the spatial and temporal information of the network output is connected using a fusion layer, and finally, after passing through a fully connected network (FCN) with a softmax layer, the final decision is made. This model has reached 90% and 73% recognition rates on EMODB and IEMOCAP datasets. Although it distinguishes the correlated emotions (*bore*, *sad*) and (*disgust*, *fear*) with high recognition rates (92.7%, 95.2%) and (90.9%, 91.3%) respectively, it still failed to distinguish (*anger*, *happiness*) well (98.1%, 66.67%).

In a study presented by [9], a deep CNN network was used to extract emotional features. Then, a correlation-based method was used to select the most appropriate and discriminative features for SER, and finally, methods such as support vector machine (SVM), random forests (RF), and neural networks were used as classifiers. In this study, using fast Fourier transform, the input signal spectrum of emotional data is used to detect the output emotional class. The results of this study on the EMODB and SAVEE datasets were 90.5% and 66.9%, respectively. Also, this paper has been able to distinguish the correlated emotions (*bore*, *sad*), (*fear*, *disgust*), and (*happiness*, *anger*) with almost high rates (87.65%, 93.54%), (92.06%, 98.91%), and (92.12%, 92.3%). But it did not recognize the *neutral* emotion well (89.87%).

Mehmet Bilal Er [10] has increased the accuracy of classification by using deep and acoustic features. First, acoustic features such as root mean square (RMS) energy and Mel frequency cepstral coefficients are extracted from the input signal. The main audio signal spectrum is then given as the input of the DNN and deep features are extracted. Then a hybrid feature vector is created and after selecting features with ReliefF algorithm and SVM classifier, the SER system will be designed. In this paper, the recognition rate on the EMODB was 90.21%. This study also distinguished correlated emotion pairs at almost high rates, (*bore, sad*) with (92%, 89%), (*fear, disgust*) with (96%, 89%) and (*anger, happiness*) with (91%, 88%), but it has recognized *neutral* emotion at a lower rate of the other references (88%).

The study presented by [11] did not use the speech acoustic features but used a residual CNN to classify speech emotions. In this model, the speaker gender information is also added to the proposed algorithm to increase the recognition rate. In this study, the emotion recognition rate on the EMODB, was 90.3%, but it did not distinguish correlated emotions with a high recognition rate. In this paper, emotions (*bore*, *sad*) with rates (100%, 80%), (*fear, disgust*) with rates (77.78%, 100%), and (*happiness*, *anger*) with rates (84.62%, 90%) have been recognized. In fact, the proposed method in this study has a classification error in distinguishing between similar emotions.

The method proposed by [12] has used a new framework for recognizing speech emotions using key sequence segment selection based on the similarity of radial base function (RBF) networks in different clusters. Then, a CNN and an RBF network were used to extract the discriminative and important features, and a long short-term memory (LSTM) network was used to classify them. Finally, the emotion recognition rates on the EMODB and IEMOCAP datasets were reported to be 85.57% and 72.25%, respectively. This paper also recognizes the correlated emotions (*bore, sad*) and (*fear, disgust*) with high rates (90%, 88%) and (87%, 92%), but it could not distinguish (*happiness*, *anger*) at high rate (66%, 91%). Also, in the research presented by [13] to solve the problem of lack of data in the emotional dataset, a module based on simple deep learning has been used.

In addition to the above methods, which use deep learning in the recognition of speech emotions, many methods have done this using generative adversarial networks (GAN). For example, a semi-supervised GAN has been used by [14] to extract the features of both labeled and unlabeled datasets. Yi and Mak [15], used an adversarial network and an auto-encoder (AE) as a classifier of emotional features. This model has a recognition rate of 84.49% and 66.92% on the EMODB and IEMOCAP datasets. In this study, the recognition rate of correlated emotions has not been reported.

Also, an adversarial AE has been presented by [16] to solve the problem of lack of emotional data which reported a recognition rate 64.5% on the IEMOCAP dataset.

Meanwhile, some studies have used wavelet transform to better recognize speech emotion. For example, Wang et al. [17], used wavelet analysis to extract features and achieved a recognition rate of 79.2% on the EMODB dataset. It has identified the pair (*anger, happiness*) with high rates (93.7%, 98.6%), while it can't distinguish well other related emotions (*bore, sad*) and (*fear, disgust*), ((75.3%, 96.8%) and (55.07%, 80.4%)). This method has also been used by [18], and its result on the EMODB was 90.09%. In this study, correlated emotions (*fear, disgust*) and (*happiness, anger*) were distinguished at a reasonable rate (88.41%, 82.61%) and (96.06%, 85.92%), while correlated emotions (*bore, sad*) was not well distinguished (82.72%, 95.16%).

Reddy and Vijayarajan [19] also used Mel-frequency cepstral coefficient and discrete wavelet transform as features of the SER system and examined their performance using the SVM and K-nearest neighbor (KNN) classifiers. Ali Bakhshi et al. [20], have used cepstral features along with a binary hidden Markov model (HMM) classifier to detect speech emotions. The recognition rate of the model presented on EMODB was 87.29%. In addition, this model distinguishes correlated emotions (*fear, disgust*) with a high rate (85.29%, 91.3%) while it has identified (*happiness*, *anger*) and (*bore, sad*) at lower rates (70.42%, 88.19%) and (100%, 88.89%). Sugan et al., [21] also used the features extracted using the triangular filter bank and the equivalent cepstral coefficients to recognize emotions. The recognition rate of the proposed system is 77.08% on EMODB. Deng Deng Li et al., (2021) also used a weighting method (inspired by feature selection methods such as minimum redundancy maximum relevance (mRMR) and ReliefF) to find emotional related features and deep learning to classify them. The emotion recognition rate of this method on EMODB was 72.19%.

In the architecture proposed by [23], high-dimensional vector of emotional features including fundamental frequency, zero cross rate, Mel-frequency cepstral coefficient (MFCC), energy and harmonic noise ratio using the c-means fuzzy clustering algorithm (FCM), are divided into different subsections and are classified using multiple RF algorithm. The recognition rate of the proposed model on the EMODB set was 85.61%. Also, this method could not distinguish well between correlated emotions (*bore*, *sad*), (*fear*, *disgust*) and (*happiness*, *anger*).

In the model proposed by [24], a new regression method called robust discriminative sparse regression is used to select appropriate discriminative features. The results on EMODB was 86.19% and only the pairs (*bore*, *sad*) could not be distinguished well.

The model proposed by [25] compares the performance of two different classifications in SER systems. In [26], an ensemble learning model random forest algorithm has been used to find the importance of different features . In this method, the weighted binary cuckoo algorithm is used to select the features and the decision tree, RF and SVM classifiers are used for classification. It has finally a recognition rate of 83.7% on EMODB. This method also distinguishes correlated emotions (*bore*, *sad*) at a better rate than (*fear*, *disgust*) and (*happiness*, *anger*). In the research presented by [27], a discriminative non-negative matrix factorization (DSNMF) method has been used to reduce the dimensions of input features. This method reached 82.8% on EMODB but could not differentiate the correlated emotions (*happiness, anger*) well. In the model proposed by [28] the Hilbert–Huang–Hurst coefficient vector (HHHC) has been used as one of the nonlinear features of the vocal source, due to its effect on speech production mechanism, to better display emotional states. In this study, gaussian mixture model (GMM), HMM, DNN, CNN and convolutional recurrent neural network (CRNN) have also been used as classifiers. This model achieved recognition rate of 81.8% on EMODB and it could not distinguish any of the correlated emotions too. And finally in the model proposed by [29] an adaptive domain-aware representation learning method is used to better identify and extract domain-dependent features. This model has a rate of 73.02% on IEMOCAP.

Table 1 summarizes the recent proposed SER methods. All of them evaluated with both criteria, weighted average recall (WAR) and unweighted average recall (UAR).

Table 1. Current researches to solve the SER system challenges

| Reference | Year | Proposed method | WAR | UAR |
|---|---|---|---|---|
| [7] | 2020 | DCNN as classifier | 86.10 | N/A |
| [8] | 2020 | CNN as classifier | N/A | 90.01 |
| [10] | 2020 | DL to extract features | 90.21 | N/A |
| [11] | 2020 | RCNN as classifier | 90.30 | N/A |
| [12] | 2020 | CNN for feature extraction and bidirectional deep neural network LSTM for feature classification | 85.57 | N/A |
| [13] | 2021 | DL to extract features | N/A | 90.01 |
| [9] | 2020 | DCNN for feature extraction | 90.5 | N/A |
| [14] | 2020 | semi-supervised GAN as classifier | 65.20 | 68.00 |
| [15] | 2020 | GAN and AE as classifier | 84.49 | 83.31 |
| [16] | 2020 | adversarial AE for features discriminative recognition | N/A | 66.70 |
| [17] | 2020 | Wavelet analysis to extract features | N/A | 79.20 |
| [18] | 2021 | Tunable Q wavelet transform to extract features | 90.09 | N/A |
| [20] | 2020 | BHMM as classifier | 87.29 | N/A |
| [21] | 2020 | Feature extraction using a triangular filter bank | 77.08 | N/A |
| [28] | 2020 | Use HHH coefficients to extract features | 81.80 | N/A |
| [29] | 2020 | Use adaptive domain-aware representation learning method to extract domain-dependent features | 73.02 | 65.86 |
| [13] | 2021 | Use sequential learning to find dependencies between features | N/A | 90.01 |
| [30] | 2020 | Use the active feature selection method to select the feature | N/A | 76.90 |
| [22] | 2021 | Use the feature weighting method based on emotional groups to select the features | 72.19 | N/A |
| [23] | 2020 | CMFC to select features | 85.61 | N/A |
| [24] | 2020 | Use of robust discriminative sparse regression to select discriminative features | N/A | 86.19 |
| [26] | 2021 | Using RF with group learning model and weighted binary cuckoo algorithm to select superior features | 83.70 | N/A |
| [27] | 2020 | Using discriminative non-negative matrix factorization to reduce features dimensions | 82.80 | 83.30 |

Besides, there have been a large number of research studies into the area of ELMs [31-35]. For example, [36] used a self-adaptive evolutionary ELM for optimizing network hidden node parameters, [37] developed an online sequential learning algorithm for single-hidden layer feedforward neural networks (SLFNs), [38] introduced multi-layer ELM which could learn feature representations using ELM-based auto-encoder. In [39], a new learning algorithm, called bidirectional ELM, had been proposed, in which the number of hidden nodes could be further reduced without affecting learning effectiveness, and finally [40] extended ELMs for both semi-supervised and unsupervised tasks based on manifold regularization.

So far, many weighting algorithms have been proposed for imbalanced data classification with ELMs including [41], in which a weighted ELM had been proposed to deal with data with imbalanced class distribution, [42] that presented an effective method based on a fast ELM classifier and fuzzy membership of samples for class-imbalanced learning, and [43] wherein previous weighted ELMs could be improved with decay-weight matrix setting for balanced and optimized learning.

## 3- System Description

As illustrated in Fig. 2, the overall SER system consists of three main parts: preprocessing and feature extraction, feature selection, and feature classification. In this section, the SER system designed using the proposed method will be elaborated.

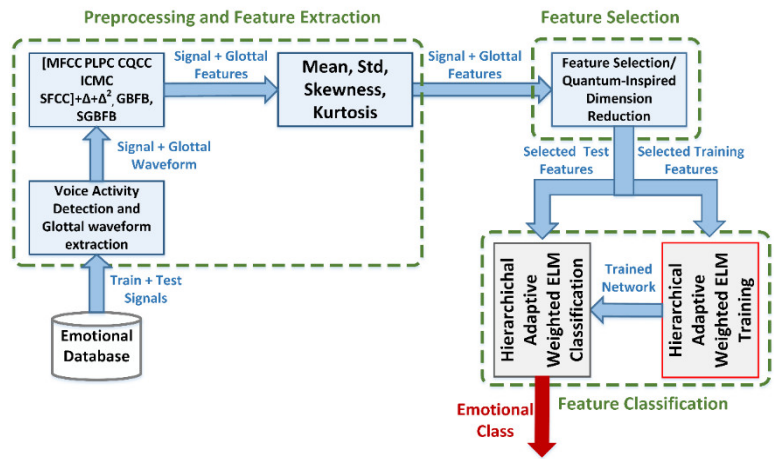

Fig. 2 Overall proposed SER system

## 3-1 Feature Extraction

It is an important stage of an efficient SER system. At first, the preprocessing of the speech signal before feature extraction has been considered. Pre-emphasizing, framing, windowing, and voice activity detection are three common techniques used in signal preprocessing. The preprocessed signal will be then fed to the feature extraction module. In this stage, the necessary and emotion-relevant features like the spectro-temporal features will be extracted from the signal, which will be explained below.

## 3-1-1 GBFB Features

There is an uncertainty relation between a signal's specificity in time and frequency. Many approaches show that both spectral and temporal information of a spectro-temporal representation of a speech signal is useful for speech processing and limiting redundant features [44]. Dennis Gabor defined a family of signals (called Gabor filters) to optimize spectral and temporal trade-off. The Gabor elementary function is the sum of Gaussian cosine and Gaussian sine functions that can minimize the space (time)-uncertainty product [1]. The Gabor filter *h(t)* is a Gaussian function modulated with a sinusoidal function with frequency *u₀* as below (Fig. 3),

$$h(t) = s(t)g(t) \tag{1}$$

Where *s(t)* shows a complex sinusoidal function:

$$s(t) = \exp(j2\pi u_0 t) \tag{2}$$

And, *g(t)* refers to a Gaussian function, also known as envelope:

$$g(t) = \frac{1}{\sqrt{2\pi}\sigma_x} \exp\left(\frac{-t^2}{2\sigma_x^2}\right) \tag{3}$$

Thus, the Gabor filter can be written as:

$$h(t) = \frac{1}{\sqrt{2\pi}\sigma_x} \exp\left(\frac{-t^2}{2\sigma_x^2}\right) \exp(j2\pi u_0 t) \tag{4}$$

And, the frequency response of the filter is:

$$H(u) = G(u - u_0) \tag{5}$$

$$H(u) = 2\pi\sigma_x \exp(-2\pi^2(u-u_0)^2\sigma_x^2) = \frac{1}{2\pi\sigma_u}\exp\left(-\frac{1}{2}\frac{(u-u_0)^2}{\sigma_u^2}\right), \quad \sigma_u = \frac{1}{2\pi\sigma_x} \tag{6}$$

*u₀* indicates the Gabor filter spatial center frequency and *σₓ* represents the standard deviation (SD) of the Gaussian envelope and determines the filter bandwidth.

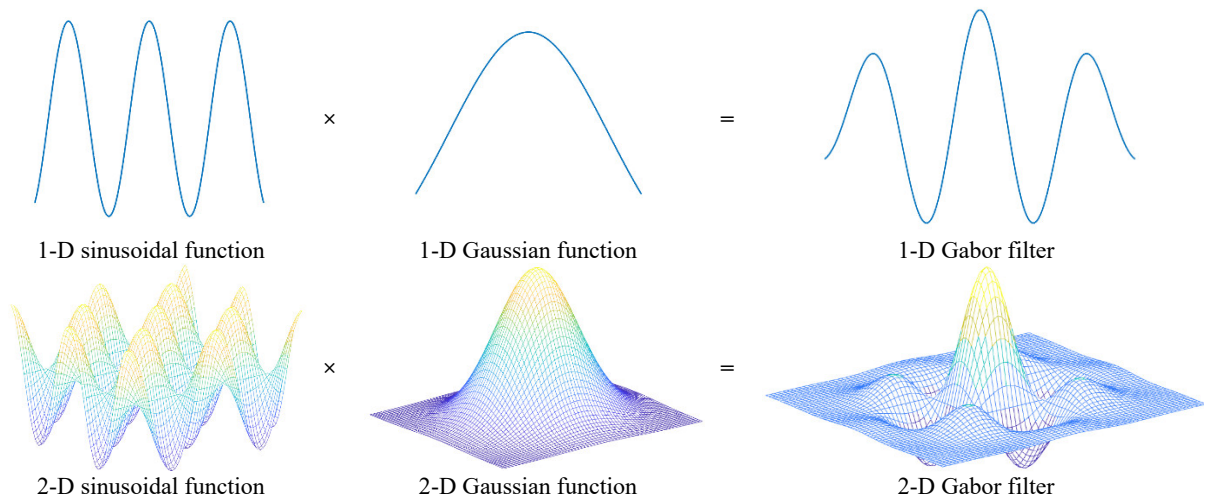

Fig. 3 Gabor filters

However, a Gabor filter in a 2D space is a sinusoidal plane of particular frequency and orientation, modulated by a Gaussian envelope (Fig. 3). It can be written as:

$$h(x,y) = s(x,y)g(x,y) \tag{7}$$

$$s(x,y) = \exp(j2\pi(u_0 x + v_0 y)) \tag{8}$$

$$g(x,y) = \frac{1}{2\pi\sigma_x\sigma_y}\exp(-\frac{1}{2}(\frac{x^2}{\sigma_x^2} + \frac{y^2}{\sigma_y^2})) \tag{9}$$

The Gabor filter can be also written as:

$$h(x,y) = \frac{1}{2\pi\sigma_x\sigma_y}\exp(-\frac{1}{2}(\frac{x^2}{\sigma_x^2} + \frac{y^2}{\sigma_y^2}))\exp(j2\pi(u_0 x + v_0 y)) \tag{10}$$

Therefore, the frequency response of the filter is:

$$H(u,v) = G(u - u_0, v - v_0) \tag{11}$$

$$H(u,v) = 2\pi\sigma_x\sigma_y \exp(-2\pi^2(\sigma_x^2(u-u_0)^2 + \sigma_y^2(v-v_0)^2)) = \\ \frac{1}{2\pi\sigma_u\sigma_v}\exp(-\frac{1}{2}(\frac{(u-u_0)^2}{\sigma_u^2} + \frac{(v-v_0)^2}{\sigma_v^2})), \quad \sigma_u = \frac{1}{2\pi\sigma_x}, \sigma_v = \frac{1}{2\pi\sigma_y} \tag{12}$$

*u₀* and *v₀* denote Gabor filter spatial center frequency. *σₓ* and *σᵧ* represent the SD of the Gaussian envelope along *x* and *y* directions and determine the filter bandwidth. An example of a two-dimensional (2D) Gabor filter is shown in Fig. 4.

Because of the locality property, the Gabor representation is physically more realistic than the Fourier one. It characterizes a band-limited signal of finite duration; in contrast, the Fourier representation of such a signal requires an infinite superposition of nonlocal signals. Also, the spectro-temporal features derived from the GBFB are endowed with good discriminative properties in both spectral and temporal domains and are also considered as a powerful tool for improving robustness, and augmenting performance of SER systems, and limiting redundancy between their features.

*Calculation of GBFB Features*

The process of GBFB feature extraction is illustrated in Fig. 5. First, a logarithmically-scaled mel-spectrogram, *i.e.* with logarithmic amplitude and mel-frequency scaling or any similar spectro-temporal representation is calculated from the speech (Fig. 5a). Then, 41 Gabor filters (Fig. 5b) are generated for all pairs of spectral and temporal modulation frequencies except for the redundant ones. After that, the log mel-spectrogram (LMspec) will be filtered with GBFB filters (convolved with each of the 41 Gabor filters) to model the response and to represent the spectro-temporal patterns.

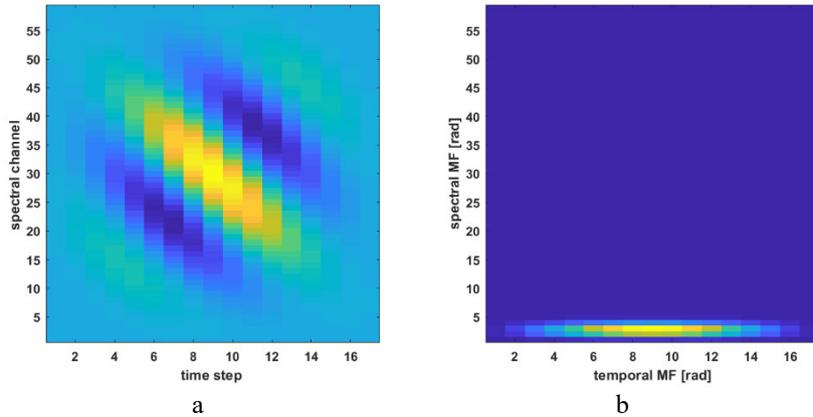

Fig. 4 a) Real part of a two-dimensional (2D) Gabor filter, b) Absolute values of filter's transfer function

However, the resulting feature vector is relatively high-dimensional when all the filter outputs of 41 filters are used. Subsequently, the filtered LMspec are spectrally sub-sampled at a rate of a quarter of the extent of the spectral width of the corresponding filter because the filter output between adjacent channels is highly correlated once the filter has a large spectral extent. This can reduce redundancy from the filtered LMspec. Finally, the filtered and sub-sampled LMspecs are also concatenated and form a 455-dimensional feature vector, referred to as GBFB features (Fig. 5c) [1].

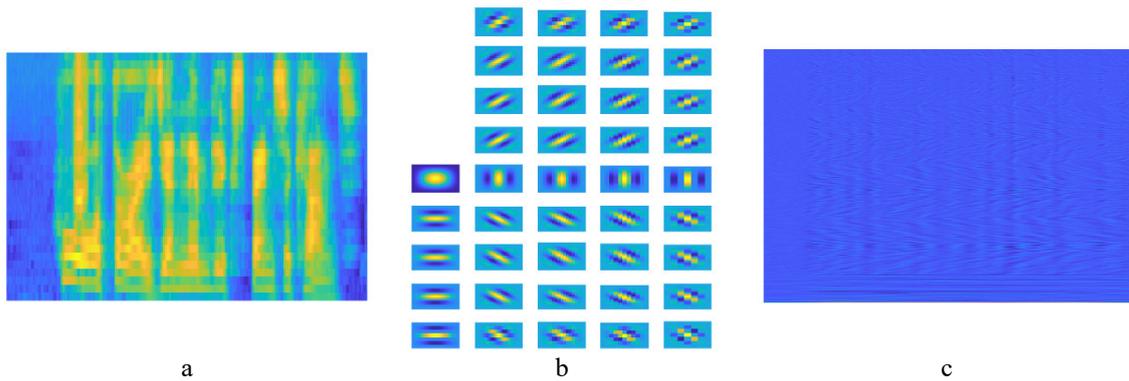

Fig. 5 Feature extraction scheme with GBFB, a) Input log mel-spectrogram b) 41 filters of GBFB, c) Resulting 455-dimensional feature vector

### *Calculation of SGBFB Features*

It has been confirmed in [1] that if a separate spectral and temporal processing with two 1D GBFB can be used to extract features, it will improve robustness in contrast to complex joint 2D spectro-temporal GBFB approach. On the other hand, the 2D GBFB can be decomposed into a spectral 1D GBFB and a temporal 1D GBFB such that the inseparable 2D patterns of the GBFB will be replaced with separable ones (a 1D Gabor filter has been illustrated in Fig. 6). SGBFB features are also those extracted with these two 1D GBFBs, instead of 2D Gabor filters. An example of an input LMspec after filtering with spectral filter, temporal filter, and spectro-temporal filter has been displayed in Fig. 7.

The experiment results in [1] revealed that since spectral and temporal processing can be performed independently in SGBFB, then it is more robust to noise compared with GBFB features.

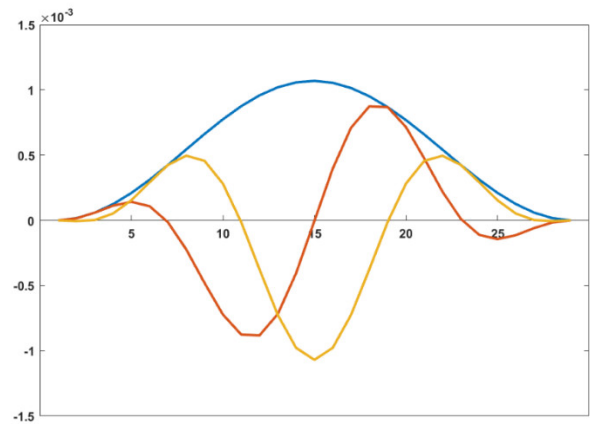

Fig. 6 1D Gabor filter, absolute (blue), real (red), and imaginary (yellow) parts

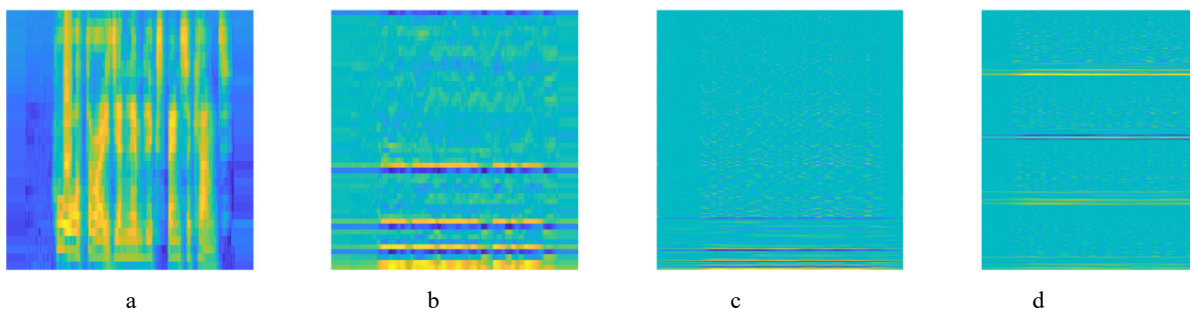

Fig. 7 Feature extraction scheme with 1D GBFB, a) Input LMspec, b) 2D convolved with spectral 1D GBFB, c) 2D convolved with temporal 1D GBFB, and d) 2D convolved with spectro-temporal 2D filter

### 3-1-2 Statistical Functions

After all the features have been extracted from the input speech signal and its glottal waveform, a feature matrix, whose rows and columns are respectively frames and different feature vector elements, will be obtained. Considering that the input signal extracted features are per frame, the evaluation of a system for SER is usually based on the whole emotional utterance. For this purpose and in order to facilitate the comparison of recognition rates, statistical data are used to find the features of related utterances.

### 3-2 Feature Selection/Dimension Reduction

The purpose of feature selection and dimension reduction methods is to reduce the dimension of the feature space to select more influential features and to reduce redundancy. So, today, these methods have become an integral component of the learning process in order to deal with high-dimensional data. Besides, a proper feature selection/dimension reduction method can improve the learning process in a variety of ways including learning speed and generalization capacity. In this experiment, due to the high dimensions of the extracted features and in order to reduce redundancy, some feature selection/dimension reduction methods will be used. They are as follows,

- mRMR [45] denotes a feature selection method for selecting features according to the mutual information in a way that there is minimum redundancy between selected features and maximum relevance between such features and their targets.
- Correlation-based feature selection (CFS) [46] ranks features based on correlations. Good feature subsets contain features, which are correlated with the class label, but uncorrelated to each other.
- Local learning-based clustering (LLC) [47] is a local learning approach for feature clustering. The basic idea is that a good clustering result should have the property that the cluster label of each feature point can be well predicted based on its neighboring features and their cluster labels.

### 3-2-1 pQPSO

So far, many changes have been made to the standard QPSO algorithm, such as the weighted QPSO (wQPSO) algorithm [48]. Point mass function-weighted QPSO (called pQPSO) is another modified version of the standard QPSO, which has been proposed by the authors in this paper [5,6].

In this modified-QPSO, mean best and global best position in the standard QPSO, is replaced by a randomly-selected particle, Lucky Global (LG) best, which is selected from a set of top $K$ best particles found in each iteration according to (13) and is then generated with point mass function distribution $p(LG^t)$ according to (14) to (16).

$$topK^t = \{topK_1^t, topK_2^t,, topK_K^t\} \tag{13}$$

$$c_k^t = 1 + \frac{E_{\max}^t - E(topK_k^t)}{E_{\max}^t - E_{\min}^t} \tag{14}$$

$$r_k^t = \frac{c_k^t}{\sum_{i=1}^{K} c_i^t} \tag{15}$$

$$p(LG^t) = \sum_{k=1}^{K} r_k^t \delta(LG^t - topK_k^t) \tag{16}$$

$$LG^t \sim p(LG^t) \tag{17}$$

In (14), $E_{max}^t$ and $E_{min}^t$ represent highest and lowest values of cost function for the $topK^t$ list members, and $r_k^t$ is particles' relative fitness factor in $topK^t$ list. After generating $LG^t$ with probability $p(LG^t)$, this value replaces both $P_g^t$ and $m_{best}^t$.

In the standard QPSO algorithm, new particles may be created outside the desired interval. Since these particles represent invalid solutions to the problem, they should be eliminated or moved to a valid range (that is, repaired). Removing or repairing the new particles also slows down algorithm convergence. To cope with this problem in the proposed pQPSO, truncated Laplace distribution (TLD) is used for generating new particle positions within the valid range. Assuming that the random variable $x$ is in the interval $B_l \leq x \leq B_u$, with mean $\mu$ and standard deviation $\sigma$, in the TLD; the probability density function $f_t$(PDF) and cumulative distribution function $F_t$(CDF) of $x$ are as follows.

$$A = 2 - e^{-\frac{B_u - \mu}{\sigma}} - e^{\frac{B_l - \mu}{\sigma}} \tag{18}$$

$$f_t(x; \mu, \sigma, B_l, B_u) = \frac{1}{A} \begin{cases} \frac{1}{\sigma} e^{-\frac{|x-\mu|}{\sigma}}, & B_l \leq x \leq B_u \\ 0, & \text{otherwise} \end{cases} \tag{19}$$

$$F_t(x) = \frac{1}{A} \begin{cases} e^{\frac{x-\mu}{\sigma}} - e^{\frac{B_l - \mu}{\sigma}}, & B_l \leq x \leq \mu \\ 2 - e^{-\frac{x-\mu}{\sigma}} - e^{\frac{B_l - \mu}{\sigma}}, & \mu \leq x \leq B_u \end{cases} \tag{20}$$

Then, particle position is generated randomly using the following equations.

$$S = \text{sgn}(F_t(\mu) - u), \quad u \sim U(0,1) \tag{21}$$

$$x = \mu + S\sigma \ln(1 + S(Au + e^{\frac{B_l - \mu}{\sigma}} - 1)) \tag{22}$$

wherein, sgn stands for the sign function.

Moreover, in this paper, an adaptive technique is proposed for determining contraction expansion value in each iteration instead of using constant or linearly-varying method for contraction expansion (CE) coefficient; in which, its value is calculated proportional to error reduction rate in the previous iteration, as follows.

$$\alpha = \left(1 - \frac{mt}{\max Try}\right) \cdot \left(\alpha_1 + (\alpha_0 - \alpha_1)\frac{(T-t)}{T}\right), \quad 0 \leq t \leq T \tag{23}$$

in which,

$\alpha_0$: is initial value,
$\alpha_1$: refers to final value of CE coefficient,
$T$: represents number of iterations,
$t$: shows current iteration,
$maxTry$: is maximum number of unsuccessful attempts to improve cost function more than a specified value $\epsilon$ relatively,
$mt$: indicates current number of unsuccessful attempts.
If the algorithm fails to reduce relative error by more than $\epsilon$ in $maxTry$ successive iterations, it is terminated.

### 3-3 Classification
Following feature selection, a classification method will be utilized to organize the selected features. This method will be trained on selected training features and then it is ready for classification of test features. The final accuracy rate on the test set actually indicates the performance of the proposed algorithm. In this section, the unweighted ELM and the proposed weighted ELM have been described.

### 3-3-1 Unweighted ELM
ELMs are random SLFNs whose purpose is to solve the problem of slow-learning back-propagation (BP) method in neural networks (NNs). In these machines, the weights between the input layer and the hidden layer are randomly assigned. Then, unlike the BP method, the weights between the hidden layer and the output layer are obtained using the regularized least squares (LS) technique to resolve the problem of minimizing the training error and the norm of output weights [49].

For example, in an ELM network having $L$ nodes in the hidden layer as well as $N$ different training pairs $(\mathbf{x}_i, \mathbf{t}_i)$, the output of the hidden layer for input vector $x_i$ can be represented as follows:

$$h(x_i) = [h_1(x_i),..,h_L(x_i)]_{1 \times L} \qquad i = 1,2,..,N \tag{24}$$

So that,

$$x_i = [x_{i1}, x_{i2},..., x_{iD}]^t \qquad i = 1,2,..,N \tag{25}$$

$$t_i = [t_{i1}, t_{i2},..., t_{iM}]_{1 \times M} \qquad i = 1,2,..,N \tag{26}$$

$M$ refers to number of classes. Then, a simple ELM network can be represented by the following relations:

$$H\beta = T \tag{27}$$

$$H = \begin{bmatrix} h(x_1) \\ \vdots \\ h(x_N) \end{bmatrix}_{N \times L} \tag{28}$$

$$\beta = [\beta_1, \beta_2,...,\beta_M]_{L \times M} \tag{29}$$

$$T = \begin{bmatrix} t_1 \\ t_2 \\ \vdots \\ t_N \end{bmatrix}_{N \times M} \tag{30}$$

The output vector $y_i$ for input $x_i$ is calculated as follows:

$$y_i = h(x_i)\beta \tag{31}$$

$$Y = \begin{bmatrix} y_1 \\ y_2 \\ \vdots \\ y_N \end{bmatrix}_{N \times M} \tag{32}$$

$$Y = H\beta \tag{33}$$

The matrix $Y$ is horizontally-stacked output vectors of the ELM for all training samples. As well, the matrix $H$ is horizontally-stacked output vectors of the hidden layer for all training samples, the matrix $T$ is horizontally-stacked target vectors for all training samples (also known as indicator matrix or one-hot matrix), and the matrix $\beta$ is vertically-stacked weight vectors of the output layer for all classes.

As stated in Bartlett's theory [50], the purpose of ELM is to reduce training error $\|H\beta - T\|_F^2$ (the Frobenius norm of error matrix) and to maximize marginal distance between classes, or to minimize the norm of output weights $\|\beta\|^2$. Accordingly, the classification problem is mathematically written as:

$$\min_{\beta}\{\frac{1}{2}\|\beta\|_F^2 + \frac{1}{2}C\|H\beta - T\|_F^2\} \equiv \min_{\beta}\{\frac{1}{2}\|\beta\|_F^2 + \frac{1}{2}C\sum_{i=1}^{N}\|\xi_i\|^2\} \tag{34}$$

$$\text{s.t.} \quad \xi_i = h(x_i)\beta - t_i, \quad i = 1,...,N \tag{35}$$

$$\xi_i = [\xi_{i1}, \xi_{i2},...\xi_{iM}] \tag{36}$$

Here, $\xi_i$ is the error vector for training sample $x_i$. So, $\beta$ is calculated in the ELM as follows:

$$\beta = H^\dagger T \tag{37}$$

$$\beta = \begin{cases} H^t(\frac{I}{C} + HH^t)^{-1}T & N < L \\ (\frac{I}{C} + H^tH)^{-1}H^tT & N > L \end{cases} \tag{38}$$

Such that $H^\dagger$ is Moore-Penrose (generalized) matrix inverse of $H$ and $C$ refers to regularization parameter for better generalization [49].

### 3-3-2 Weighted ELM
*Previous Work*

Suppose the data distribution has two classes of majorities and minorities; where the majority class sample numbers are more than average and the minorities are below average. If the unweighted ELM method is used, the samples in the majority and minority classes will have the same weight. Therefore, the majority class boundary will be increased to reduce the training error and to have the same misclassification costs for each sample, and some examples of the minority classes will be then mistakenly classified as the majority ones. The aim of weighted ELM design is to moderate the majority class boundary, to enlarge the boundary of the minority one, and consequently to diminish the training error. Then, the weight of the samples is used in ELM training to balance data distribution and (34) is changed into:

$$\min_{\beta}\left\{\frac{1}{2}\|\beta\|_F^2 + \frac{1}{2}C\left\|W^{\frac{1}{2}}(H\beta - T)\right\|_F^2\right\} = \min_{\beta}\left\{\frac{1}{2}\|\beta\|_F^2 + \frac{1}{2}C\sum_{i=1}^{N}W_{ii}\|\xi_i\|^2\right\} \tag{39}$$

$$W = \mathrm{diag}(W_{11}, W_{22}, ..., W_{NN}) \tag{40}$$

Where $W$ is an $N \times N$ diagonal positive matrix and $W_{ii}$ corresponds to training sample $x_i$. The purpose of this type of ELM is, if $x_i$ belongs to a minority class, its weight will be greater than that of the majority class to balance the distribution of data. In this case, the relation (38) will change as follows:

$$\beta = \begin{cases} H^t W(\frac{I}{C} + HH^tW)^{-1}T = H^t(\frac{I}{C} + WHH^t)^{-1}WT & N < L \\ (\frac{I}{C} + H^tWH)^{-1}H^tWT & N > L \end{cases} \tag{41}$$

So far, several algorithms have been used as weighting methods to balance the class boundaries. The following two weighting methods are proposed in [41]:

$$\mathbf{W1}: \quad W_{ii} = \frac{1}{N_{c(i)}} \quad i = 1, ..., N \tag{42}$$

$$\mathbf{W2}: \quad \begin{cases} W_{ii} = \frac{0.618}{N_{c(i)}} & N_{c(i)} > \mathbf{AVG} \\ W_{ii} = \frac{1}{N_{c(i)}} & N_{c(i)} \leq \mathbf{AVG} \end{cases} \tag{43}$$

Where $c(i) \in \{1, 2, ..., M\}$ is class of sample $x_i$ and $N_c$ denotes number of samples belonging to class $c$ and AVG represents average number of samples in all classes.

With this weighting method, the greater the number of samples in a class, the lower the weight of those samples will be, and vice versa. In the second weighting method, this ratio is further reduced and the samples from the majority classes will gain less weight than the first method.

One of the problems with this weighting method is that, the weight of the samples of the majority classes decreases relative to the number of samples; then, some of the samples from the majority classes will be misclassified as minority ones. This problem will be more severe in the second method because the weight of the majority classes is further decreased.

To explain this problem, a new method has been proposed in [42], which has the following weighting relation in multi-class and binary classification cases:

$$\mathbf{W3}: \quad W_{ii} = \frac{\sqrt[d]{\frac{N_{c(i)}}{\max(N_{c(i)})}}}{N_{c(i)}} \quad i = 1, ..., N \tag{44}$$

$$W_{ii} = \begin{cases} \dfrac{\sqrt[d]{\dfrac{N_1}{\max(N_1, N_2)}}}{N_1} & \text{if } c(i) = 1 \\ \dfrac{\sqrt[d]{\dfrac{N_2}{\max(N_1, N_2)}}}{N_2} & \text{if } c(i) = 2 \end{cases} \qquad (45)$$

This weighting method increases the weight of classes by a factor proportional to the number of samples in each class to avoid the incorrect classification of the samples in the majority class into the minority one. Thus, the weight of both majority and minority classes is added in a slight manner. Since this increase is proportional to the number of instances in each class, it will be higher for the majority classes than for the minority ones. As a result, the weight loss of the majority classes is lower than the W1 and W2 methods and the misclassification rate of the majority class samples is reduced compared with that of two previous methods. The coefficient $d$ is called decaying parameter, so the greater the value, the higher the importance of the minority classes.

On the other hand, in the method presented in [43], weights are obtained from the following equation:

$$\mathbf{W4}: \quad W_{ii} = \frac{N^s_{M-r(c(i))+1}}{\sum_{j=1}^{M} N_j} = \frac{N^s_{M-r(c(i))+1}}{N} \quad i = 1, \ldots, N \qquad (46)$$

Where, the number of samples in different classes i.e. $\{N_1, N_2, \ldots, N_M\}$ is sorted in an ascending order to achieve a sorted list $N^s_1 \leq N^s_2 \leq \cdots \leq N^s_M$ and $r(c)$ is rank of class $c$ in this list.

For example, in a binary classification, the weight of the two classes with negative and positive samples as majority and minority classes (i.e. $N_1 > N_2$), is obtained as follows:

$$W_{ii} : \begin{cases} \dfrac{N_2}{N_1 + N_2} & \text{if } c(i) = 1 \\ \dfrac{N_1}{N_1 + N_2} & \text{if } c(i) = 2 \end{cases} \qquad (47)$$

***Proposed Weighting Method***

As explained before, one of the problems with the W1 weighting method is that, the weight of the samples of the majority classes decreases relative to the number of samples, then some of the samples from the majority classes will be misclassified as minority ones. This problem will be more severe in W2 method because the weight of majority classes is further decreased.

In the method presented in this study, the main objective is to boost the boundary of the majority classes more than the W1 method and to reduce the boundary of the minority ones less than W1. In this weighting method, the importance of majority classes is slightly higher than the methods W1 and W2 and the misclassification will be demoted.

The proposed formula for weighting different classes in this study is as follows:

$$\mathbf{W}: \begin{cases} W_{ii} = \dfrac{1}{p_i + (N_{c(i)} - p_i) \times \dfrac{N_{c(i)}}{\max(N_{c(i)})}} \quad i = 1, \ldots, N \\ p_i = \sum_i N_{c(i)} - N_{c(i)} \end{cases} \qquad (48)$$

In a binary mode, with negative and positive samples as majority and minority classes, it is as follows:

$$W_{ii} : \begin{cases} \dfrac{1}{N_2 + (N_1 - N_2)\dfrac{N_1}{\max(N_1, N_2)}} & \text{if } c(i) = 1 \\ \dfrac{1}{N_1 + (N_2 - N_1)\dfrac{N_2}{\max(N_1, N_2)}} & \text{if } c(i) = 2 \end{cases} \qquad (49)$$

### 3-3-3 H-ELM

The H-ELM, is kind of ELM with more than one hidden layer. Each hidden layer is an auto-encoder ELM (AE-ELM). Therefore, as shown in Fig. 8, the H-ELM training consists of two successive stages, an unsupervised hierarchical feature representation and a supervised feature classification. In the first phase, an ELM random feature space exploits hidden information among training samples and an *N*-layer unsupervised-learning ELM-based sparse auto-encoder extracts multilayer sparse features of the input data. Within this phase, each hidden layer of the H-ELM is an independent module functioning as a separated feature extractor. After unsupervised hierarchical training, the outputs of the *N*-th layer are also viewed as the high-level features extracted from the input data. They are thus utilized as the inputs of the supervised ELM-based regression to obtain the final results of the whole network. On the other hand, the supervised feature classification stage is performed for final decision-making. Algorithm 1 presents the multi-layer unsupervised-learning ELM-based sparse auto-encoder pseudo-code in detail.

The proposed H-ELM structure achieves more compact and meaningful feature representations than the original ELM. Exploiting the advantages of ELM random feature mapping, the hierarchically-encoded outputs are randomly projected before final decision-making, leading to a better generalization with faster learning speed unlike the greedy layer-wise training of DL, in which the hidden layers are trained in a forward manner with unsupervised initialization and need to be retrained (that is, fine-tuned) iteratively using back-propagation (BP)-based NNs. However, in H-ELM, the weights of the current layer are fixed without fine-tuning once the previous layer is established. Thus, H-ELM training would be much faster than that of DL.

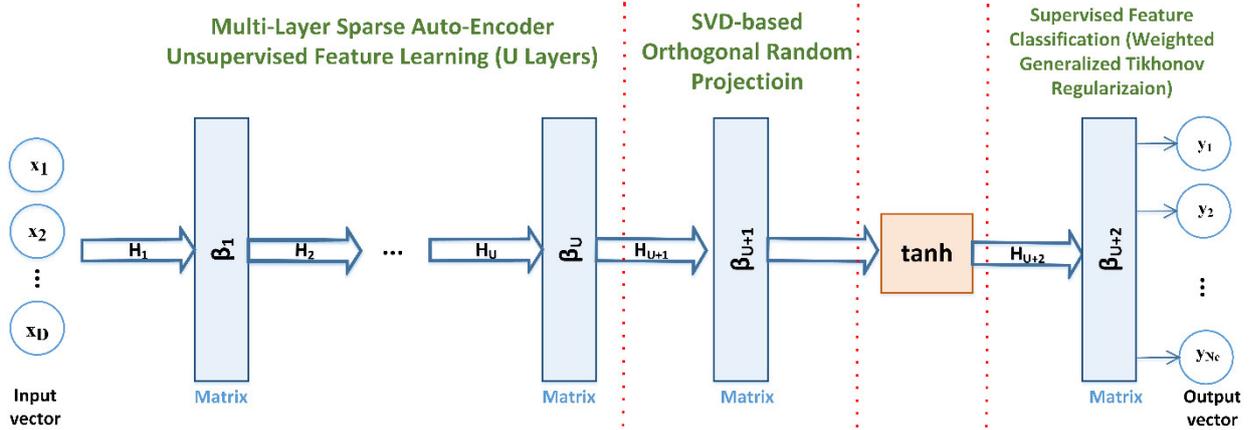

Fig. 8 H-ELM structure (details are explained in the algorithm 1)

## 4- Experiments
### 4-1 Emotional Datasets

Many emotional databases have been designed to test the performance of SER systems. In this study, three common databases including Berlin Database of Emotional Speech (EMODB) [51], Surrey Audio-Visual Expressed Emotion (SAVEE) [52], and the Interactive Emotional Dyadic Motion Capture (IEMOCAP) [53] were employed to evaluate the effectiveness of the proposed system, whose specifications are listed in Table 2.

Table 2. Specifications of emotional datasets

| Dataset | Language | Total no. of Speakers | Emotions | | | | | | | | Total |
|---|---|---|---|---|---|---|---|---|---|---|---|
| | | | Anger (A) | Disgust (D) | Fear (F) | Happiness (H) | Sad (SA) | Bore (B) | Surprise (SU) | Neutral (N) | |
| EMODB | German | 10 | 127 | 46 | 69 | 71 | 62 | 81 | - | 79 | 535 |
| SAVEE | English | 4 | 60 | 60 | 60 | 60 | 60 | - | 60 | 120 | 480 |
| IEMOCAP | English | 10 | 1103 | - | - | 1636 | 1084 | - | - | 1708 | 5531 |

### 4-2 Feature Extraction

Since the best and the most effective features for SER have not yet been identified, many experiments have been done in this study to select the best set of features, whose results are presented in Table 3. Given that such features are obtained in diverse ways, each one describes a speech signal from a different perspective. In Table 2, the number of dimensions examined for each feature as well as related references and their recognition rates (i.e.

WAR) are also illustrated. All the experiments have been conducted on EMODB corpus in a 10-fold cross-validation case (that is, leave one speaker out (LOSO)) using the SVM classifier.

---

**Algorithm 1** Pseudo-Code for Training Deep Sparse-AutoEncoder ELM Classifier

1: **Input:** $X \in \mathbb{R}^{N \times D}$ (Training Data Matrix), $T \in \mathbb{R}^{N \times N_{\text{emot}}}$ (Targets in One-Hot Matrix Form), $U$ (Number of Unsupervised Sparse Layers), $W \in \mathbb{R}_+^{N \times N}$ (Sample-Wise Importance Weights Diagonal Matrix)

2: **Output:** $\{H_i, \beta_i\}$

   // First Stage: Unsupervised Feature Learning Using Multi-Layer Sparse Auto-Encoders

3: $H_1 = X$

4: **for** $i = 1 : U$ **do**

5:      Add a Vector of 1's to the Input Matrix of the Layer (for Considering Bias) $G_i = [H_i \ \mathbf{1}]$

     Generate a Temporary Uniformly-Distributed Temporary Weight Matrix of the Layer
$$\beta_i^{\text{tmp}} \sim \mathcal{U}(-1, +1)^{(N_i+1) \times N_{i+1}}$$

6:      Compute Temporary Output of the Layer $A_i = G_i \beta_i^{\text{tmp}}$

7:      Compute the Final Sparse Weight Matrix of the Layer $\beta_i$ by Solving the following sparse least-squares problem (sparse linear inverse problem):

$$\alpha_i = \arg\min_{\alpha}\left\{\|A_i\alpha - G_i\|_2^2 + \lambda\|\alpha\|_1\right\} = \left\{\|G_i\beta_i^{\text{tmp}}\alpha - G_i\|_2^2 + \lambda\|\alpha\|_1\right\} = \left\{\|G_i(\beta_i^{\text{tmp}}\alpha - I)\|_2^2 + \lambda\|\alpha\|_1\right\}$$

8:      $\beta_i = \alpha_i^{\text{t}}$

9:      Compute Final Output of the Layer $H_{i+1} = G_i \beta_i$

     **end for**

   // Second Stage: Orthogonal Random Projection Layer Base on SVD

   Add a Vector of 1's to the Input Matrix of the Layer (for Considering Bias) $G_{U+1} = [H_{U+1} \ \mathbf{1}]$

   Generate a Temporary Uniformly-Distributed Random Weight Matrix:
   $\beta_{U+1}^{\text{tmp}} \sim \mathcal{U}(-1, +1)^{(N_U+1) \times N_{U+1}}$

   Compute Orthonormal Basis for the Range of the Matrix $\beta_{U+1}^{\text{tmp}}$ Using Compact SVD as Follows:

   **if** $N_U + 1 \geq N_{U+1}$ **then**

        $[U_1, \Sigma_1, V_1] = \text{CompactSVD}(\beta_{U+1}^{\text{tmp}})$      ( Assuming that $\beta_{U+1}^{\text{tmp}}$ is Full-Rank )

        $\beta_{U+1} = U_1$

   **else**

10:      $[U_1, \Sigma_1, V_1] = \text{CompactSVD}((\beta_{U+1}^{\text{tmp}})^{\text{t}})$      ( Assuming that $\beta_{U+1}^{\text{tmp}}$ is Full-Rank )

        $\beta_{U+1} = U_1^{\text{t}}$

   **end if**

   Compute Output of the Layer $H_{U+2} = \tanh(G_{U+1} \beta_{U+1})$

   // Third Stage: Supervised Feature Classification Layer

   Estimate Weight Matrix of Last Layer Using Weighted Generalized Tikhonov Regularization as Follows:

   **if** $N \geq N_{U+1}$ **then**

   $$\beta_{U+2} = \left(\frac{I}{C} + H_{U+2}^{\text{t}} W H_{U+2}\right)^{-1} H_{U+2}^{\text{t}} W T \quad \text{(Over-Determined Case)}$$

   **else**

   $$\beta_{U+2} = H_{U+2}^{\text{t}} W \left(\frac{I}{C} + H_{U+2} H_{U+2}^{\text{t}} W\right)^{-1} T \quad \text{(Under-Determined Case)}$$

   **end if**

Features evaluated in this study include MFCC [54], perceptual linear predictive cepstral (PLPC) [55], perceptual minimum-variance distortionless response cepstral coefficient (PMVDR) [56], pitch (F0) [54], glottal waveform signal [54], CQCC [2], ICMC [4], frequency domain linear prediction (FDLP) [57], cochlear filter cepstral coefficient (CFCC) [58], GBFB [1], SGBFB [1], wavelet cepstral coefficient (WCC) [59], and SFCC [60] along with their first- and second-order derivatives (i.e. Δ, Δ²) for the input speech signal.

Table 3. Evaluated feature extraction methods and their recognition rates (WAR) on EMODB dataset

| Feature Type | Feature dimension | WAR | Feature Type | Feature dimension | WAR |
|---|---|---|---|---|---|
| Pitch | 1 | 42.12% | MFCC + Δ | 14 | 69.66% |
| Pitch + Δ | 2 | 45.13% | MVDR | 21 | 70.03% |
| Pitch + Δ + Δ² | 3 | 46.31% | MVDR + Δ + Δ² | 63 | 70.39% |
| MFCC | 7 | 56.88% | MVDR + Δ | 42 | 72.43% |
| ICMC | 20 | 57.64% | PLPC + Δ + Δ² | 60 | 73.10% |
| PLPC | 7 | 62.87% | MFCC + Δ + Δ² | 60 | 73.31% |
| MFCC | 20 | 64.53% | CQCC + Δ + Δ² | 60 | 73.42% |
| CFCC | 16 | 64.78% | PLPC + Δ | 14 | 73.45% |
| PLPC | 20 | 66.35% | PLPC + Δ + Δ² | 21 | 74.51% |
| WCC | 40 | 67.05% | ICMC + Δ + Δ² | 60 | 76.53% |
| SFCC | 13 | 67.13% | SFCC + Δ + Δ² | 39 | 76.61% |
| MFCC + Δ + Δ² | 21 | 67.62% | GBFB + SGBFB + Δ | 2950 | 83.58% |
| CQCC | 20 | 68.33% | SGBFB | 1020 | 83.82% |
| FDLP | 39 | 68.86% | GBFB + SGBFB | 1475 | 85.26% |
| Glottal | 60 | 69.34% | GBFB | 455 | 85.50% |

Considering the impact of features evaluated, the following feature set are selected from the list of the mentioned features (the numbers in parentheses represent the number of dimensions of each feature), MFCC (20), PLPC (7), CQCC (20), ICMC (20), SFCC (13), and their first and second-order derivatives as well as GBFB (455) and SGBFB (1020) features. The extracted feature vector dimensionality is equal to 1715. In addition to the above feature vector, the same features will be extracted from the pitch-shifted speech signal (by 0.9 and 1.1) and added to the input training samples which thereby increase the accuracy of the SER system.

### 4-2-1 Statistical Functions

There are many statistical functions to extract statistical data from the frames of features. To find the best appropriate ones, some experiments have been performed in this study. Table 4 shows some statistical functions examined including their relations, where $x$ is a random variable having $n$ observations ($i = 1, …, n$). These functions and their combinations have been also applied to the extracted feature vector obtained from the previous sub-section. The recognition rates have been illustrated in Fig 9. Moreover, all the experiments have been performed on EMODB corpus using the SVM classifier in LOSO case.

According to the results, the statistical functions that reach the best recognition rate consist of mean, SD, skewness, kurtosis that can prepare the related feature vectors of utterances. Therefore, the dimension of the extracted feature vector will increase from 1715 to 6860 after applying the mentioned statistical functions.

Table 4. Statistical functions and their relations

| Statistical Function | Relation | Statistical Function | Relation |
|---|---|---|---|
| Mean | $\bar{x} = \frac{1}{n}\sum_{i=1}^{n} x_i$ | Kurtosis | $Kurt = \frac{1}{n}\sum_{i=1}^{n}(\frac{x_i - \bar{x}}{\sigma})^4$ |
| Quadratic mean | $\sqrt{\frac{1}{n}\sum_{i=1}^{n} x_i^2}$ | Percentile | $R = \lceil \frac{P \times N}{100} \rceil$, $P$ = percentile (here $P$=50), $R$ = percentile rank, $P$-th percentile = $x_R$ |
| Harmonic mean | $(\frac{1}{n}\sum_{i=1}^{n} x_i^{-1})^{-1}$ | Zero crossing (ZC) | $ZC = \frac{1}{2N}\sum_{i=2}^{N}\|\text{sgn}(x_i) - \text{sgn}(x_{i-1})\|$, $\text{sgn}(x_i) = \begin{cases} +1 & x_i \geq 0 \\ -1 & x_i < 0 \end{cases}$ |
| Standard deviation (std) | $\sigma = \sqrt{\frac{1}{n-1}\sum_{i=1}^{n}(x_i - \bar{x})^2}$ | Geometric mean | $\sqrt[n]{\prod_{i=1}^{n} x_i}$ |
| Skewness | $Skew = \frac{1}{n}\sum_{i=1}^{n}(\frac{x_i - \bar{x}}{\sigma})^3$ | | |

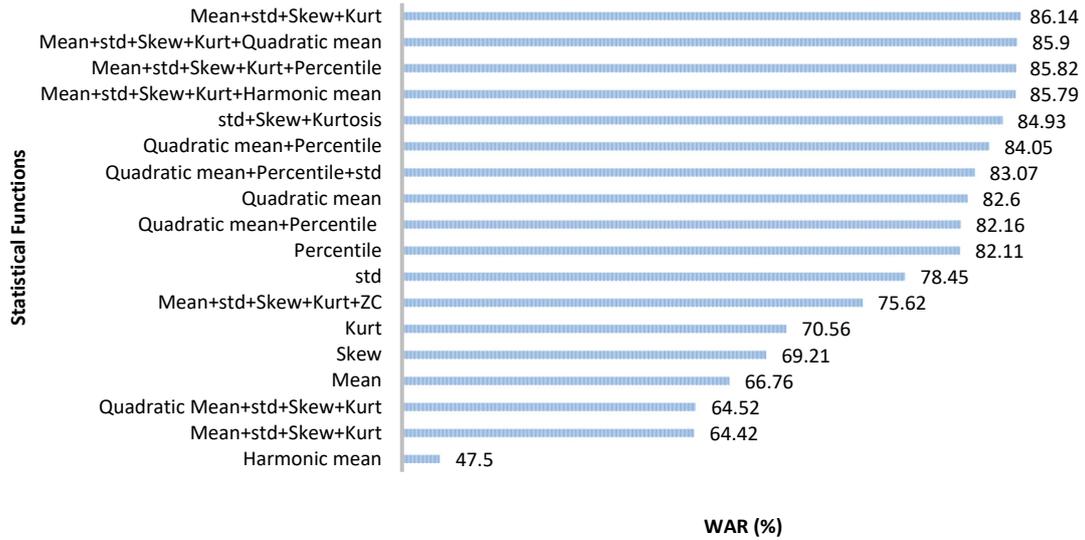

Fig. 9 Statistical functions and recognition rates

## 4-3 Feature Selection/Dimension Reduction

In this experiment, due to the high dimensions of the extracted features and in order to reduce redundancy, a feature selection method will be used. Before applying this algorithm, in order to find the best number of selected dimensions, several experiments have been performed on EMODB corpus by the SVM classifier and in LOSO case, the diagram of the cumulative sum of their scores is shown in Fig. 10.

Since the highest cumulative sum of weights is for 3000 and 5000 dimension numbers, the subsequent experiments will be then performed with both 3000 and 5000 dimensions. Then, in order to find the best feature selection/dimension reduction method, the proposed H-AWELM model will be implemented using different feature selection/dimension reduction methods.

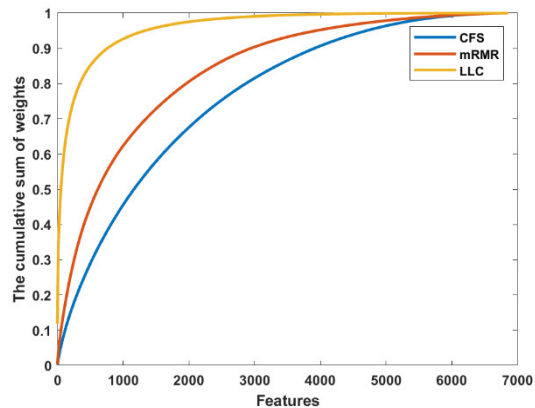

Fig. 10. The cumulative sum of features' scores for three different feature selection methods

### 4-3-1 Classical Feature Selection Methods

The related recognition rates for different feature selection methods are listed in Table 5. Among these, the mRMR method has had better results than other feature selection methods.

Table 5. The recognition rate using different feature selection methods in the H-AWELM model

| Feature selection method | Feature selection No | WAR |
|---|---|---|
| LLC | 5000 | 89.58% |
| CFS | 5000 | 90.06% |
| mRMR | 3000 | **91.29%** |
| | 5000 | **91.24%** |

### 4-3-2 Dimension Reduction Methods

In this case, the proposed model is implemented with different feature dimension reduction methods such as pQPSO, wQPSO and QPSO and the results are shown in Table 6. Meanwhile, the pQPSO method has had better results than other dimension reduction methods.

Table 6. The recognition rate using different dimension reduction methods in the H-AWELM model

| Dimension reduction method | Dimension reduction No | WAR |
|---|---|---|
| QPSO | 5000 | 88.11% |
| WQPSO | 5000 | 89.38% |
| pQPSO | 5000 | **90.04%** |

However, since the recognition rate obtained by the mRMR method was higher than pQPSO and other feature selection methods (Table 5-6), in the final experiments, the mRMR algorithm was used for feature selection.

### 4-4 Classification

In this section, in order to find the most appropriate structure for the mentioned H-ELM, some experiments have been performed on different layers and various neuron numbers (Fig 11). Based on the results presented on EMODB corpus, the best structure of unsupervised layer is belonging to a structure with three different layers with 1000, 1000, and 20000 neurons.

On the other hand, in the last layer of the H-ELM, diverse functions can be used as an ELM kernel function. The results for some functions are shown in Fig. 12. According to the results, the tangent hyperbolic transfer function has produced better recognition rates compared with other functions.

Therefore, the experiments will continue with these hidden layer neurons and kernel function including the proposed adaptive weighted algorithm in the last layer of the H-ELM.

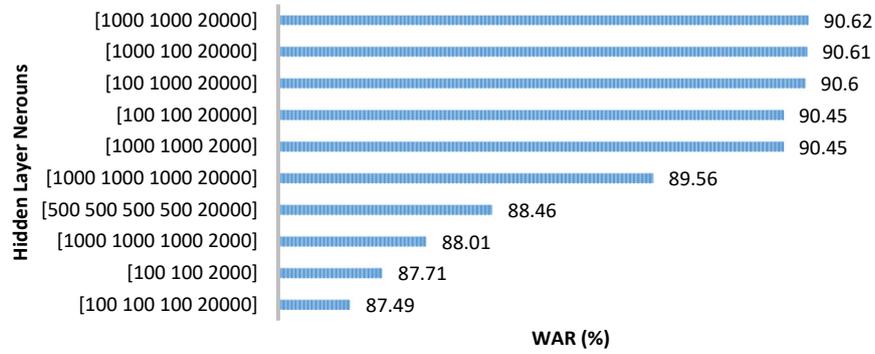

Fig. 11 H-ELM hidden layer and neuron numbers

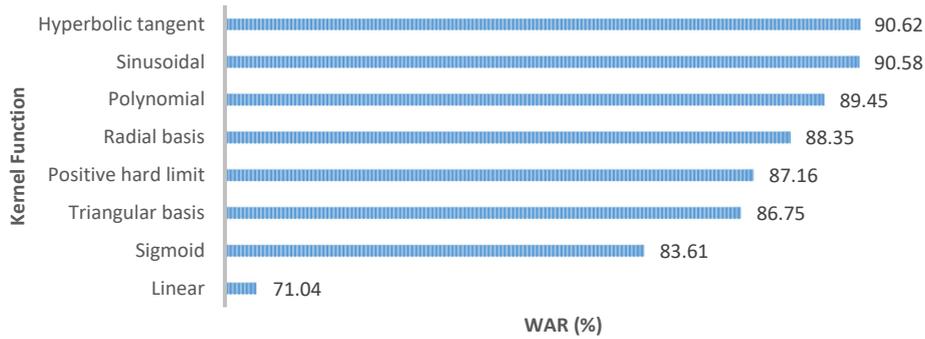

Fig. 12 Recognition rates of different kernel functions

## 5- Performance Evaluation

After the feature extraction stage of the training and test signals, a final model of the three-layer H-ELM with the tangent hyperbolic kernel and the proposed adaptive weighting approach has been learned on the selected training features to classify the selected test features. The final recognition rate on the test features actually indicates the performance of the proposed algorithm.

In this section, for better comparison of the obtained results, the criterion of imbalance ratio (IR) is used. This criterion is used to evaluate the degree of dataset imbalance and in the case of multiclass, is obtained from the following relation [43],

$$IR = \frac{\min \#(t_i)}{\max \#(t_i)}, i = 1, \dots, N \tag{50}$$

$\#(t_i)$ refers to the number of instances of class $t_i$. This ratio has been shown for different datasets in Table 7. As it turns out, all three datasets have an IR less than 1 and they suffer from a kind of imbalance. Since these datasets have different imbalance ratios, they are suitable for evaluating the performance of the proposed weighting algorithm.

Table 7. The degree of IR of different datasets

| Dataset | EMODB | SAVEE | IEMOCAP |
|---|---|---|---|
| IR | 0.3622 | 0.5 | 0.6347 |

## 5-1 Comparison of Results on Different Weighting Algorithms

The results of the proposed H-AWELM model concerning three different databases have been illustrated in Table 8. Here, different weighting algorithms (previously presented in Section 3) implemented with the H-ELM method and their results on EMODB, SAVEE and IEMOCAP datasets have been respectively presented for both 3000 and 5000 feature selection numbers for further comparisons. Since the weighted and unweighted ELM network are randomly initialized, all the presented results in this paper have been repeated 10 times with 10 different random number generator seeds and the results for 10 runs are averaged whose means and standard deviations are reported ($\mu \pm \sigma$). They are also evaluated with both criteria, WAR and UAR, for the purpose of accuracy measurement.

Table 8. Recognition rates of different weighting methods

| DB | Weighting Method | Feature Selection Number | | | |
|---|---|---|---|---|---|
| | | 5000 | | 3000 | |
| | | WAR (%) | UAR (%) | WAR (%) | UAR (%) |
| EMODB | W1 | 91.14±0.34 | 88.74±0.74 | 91.29±0.44 | 88.88±0.82 |
| | W2 | 90.23±0.3 | 88.02±0.18 | 90.59±0.57 | 88.14±0.5 |
| | W3 | 91.2±0.28 | 88.83±0.54 | 90.81±0.54 | 88.82±0.8 |
| | W4 | 90.97±0.36 | 88.84±0.48 | 90.98±0.4 | 88.85±0.74 |
| | Without Weighting | 88.81±1.71 | 86.06±1.40 | 89.74±1.67 | 86.93±1.38 |
| | Proposed Weighting | **91.24±0.32** | **88.83±0.11** | **91.29±0.43** | **88.95±0.82** |
| SAVEE | W1 | 65.5±1.18 | 67.11±1.3 | 65.75±0.92 | 67±0.83 |
| | W2 | 65±0.91 | 66.83±1.04 | 64.71±0.67 | 57.22±0.41 |
| | W3 | 64.9±0.41 | 67.67±1.32 | 65.92±0.98 | 67.11±0.9 |
| | W4 | 60.31±1.18 | 62.68±1.38 | 58.94±1.04 | 57.59±0.35 |
| | Without Weighting | 57.24±2.15 | 58.84±2.12 | 55.17±2.41 | 56.04±2.10 |
| | Proposed Weighting | **65.73±1** | **67.74±1.44** | **66.94±0.65** | **67.9±0.62** |
| IEMOCAP | W1 | 63.63±0.39 | 59.21±0.44 | 62.09±0.37 | 57.22±0.41 |
| | W2 | 52.2±0.35 | 55.04±0.38 | 50.25±0.43 | 55.04±0.38 |
| | W3 | 65.76±0.26 | 60.1±0.43 | 64.29±0.28 | 62.72±1 |
| | W4 | 65.38±0.27 | 62.72±1 | 64.39±0.18 | 62.75±21 |
| | Proposed Weighting | **67.68±0.33** | **63.13±0.43** | **64.78±0.34** | **63.01±0.14** |

As shown in Table 8 the weighting method W2 is less efficient in both feature selection modes than the others. The main reason is that, in this case, the weight given to the majority classes is less than that specified by other methods. Also, the weighting method W1 does not yield relatively good results because the number of samples in each set increases and the related weight will be moderated, so the weight given to the majority classes is reduced by the same proportion. Since this weight is higher than that obtained by the W2 method, it also yields relatively better results in both feature selection modes in contrast to W1. In this research, the overall purpose of the proposed method is to decrease the training classification error. Therefore, the proposed adaptive weighting method provides higher recognition rates than the weighting methods presented in the literature review section in both 3000 and 5000 selected features number cases. Also, in comparison with the unweighted classification method, the obtained results have shown that the use of weighted classification with any of the weighting methods has achieved better results than the unweighted classification. In other words, H-AWELM can not only improve the minority classes' classification but can also maintain the majority classes' classification at the same ELM level.

In addition, the G-mean criterion is used to better understand and evaluate the classification of unbalanced datasets. This criterion, as defined below, is the geometric mean of the recall values of all classes,

$$\text{G-mean} = (\prod_{j=1}^{M} R_j)^{\frac{1}{M}} \tag{51}$$

Where $R_j$ is the recall value of class $j$. In fact, it measures the balance/trade off ratio between different emotion recall values. For example, in a binary classification problem with 99% majority sample and 1% minority sample, an unweighted classification achieves 99% accuracy by classifying all samples in the majority class. But its G-mean value will be zero because the value of the minority class recall is zero [41, 43]. Fig. 13 shows the value of the G-mean variable for the different weighting methods. As shown in this figure, the proposed method has a better value for this criterion.

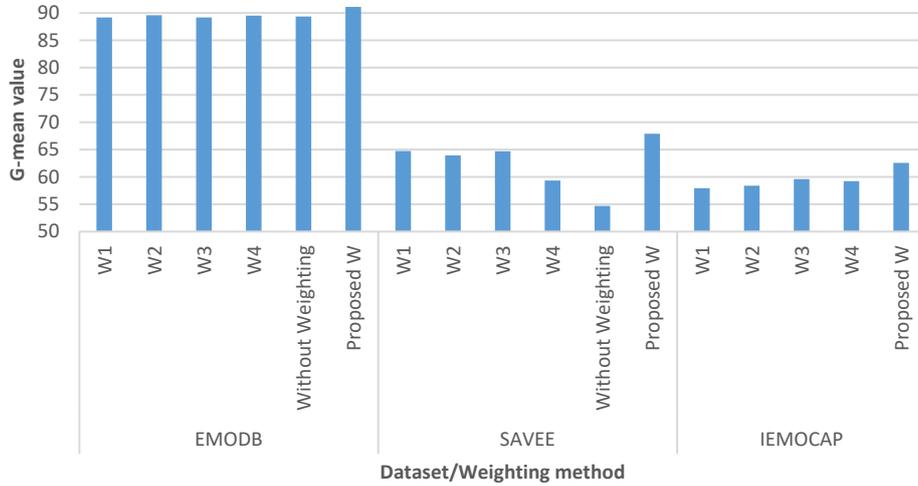

Fig. 13. The G-mean value of different weighting methods

## 5-2 Comparisons with State-of-the-Art Results

The results presented in this paper have also improved significantly compared with the findings provided by many papers in this field. Tables 9 to 11 better compare these results in terms of both WAR and UAR criteria. N/A indicates that the paper has presented its results only in WAR form.

In Table 9, both accuracy rates (WAR and UAR) of the currently published papers on EMODB corpus are compared with the results of the presented method in this paper. Obviously, the proposed algorithm has produced a higher accuracy rate than many previous methods and these results are even better than those published earlier by the authors. The recognition rates of 82.82% and 79.94% obtained in [5] and [6] were less than the 91.29% obtained in this paper. Also Fig. 14 compares the presented proposed method with other recently published methods on the EMODB dataset. In this comparison, the results of the weighted ELM classifier were better than the deep and adversarial classifiers.

In addition, Table 10 compares the results for SAVEE with the findings reported in recently published papers in terms of both WAR and UAR criteria. However, the results of the proposed system and the weighting method have been better than those in many recent papers. These results are also better than the two previous works reported by the authors. Recognition rates of 60.79% [5] and 59.38% [6] can be thus compared with 66.94% stated in this paper. Also, in Table 11, this comparison is made for the IEMOCAP corpus and recently published papers. According to the results presented in Tables 9 to 11, since the EMODB dataset has a more unbalanced distribution of data than the SAVEE and that part of the IEMOCAP examined in this study, the weighted classifier H-AWELM was more accurate in recognizing the emotions of the EMODB (91.29%) than the SAVEE (66.94%) and IEMOCAP (67.68%) datasets. In addition, the presented method is compared with the recent references in terms of *execution time* and *response time per one second of the audio file* in Fig. 15. The execution time of [5], was 0.75 hours for each EMODB fold and 7.5 hours on all 10 folds, on (Corei3-8GB RAM) processor (Fig. 15(a)). Despite the high *execution time*, it has a short *response time,* 0.6 sec (Fig. 15(b)). On the other hand, [6] has more execution time on the same processor. The use of GMM statistical classifier with high components (1024) has been one of the most important reasons for the high execution time of this algorithm. However, the time required to run the proposed H-AWELM algorithm on each fold of EMODB with the same processor specifications is 0.005 hours, which eventually reaches 0.05 hours on 10 folds (the *response time* of this model is almost zero). In addition, while H-AWELM takes less time to train the network, it also has less *response time* than other methods. Compared to other references, although the processors tested were not the same and many references used graphical processing unit (GPUs) like Yi 2020 [15], the fast ELM classifier in the H-AWELM method, has a short execution time and less response time compared to recent methods. In fact, the method presented in this paper has not only achieved better results than deep methods, but also has less execution time and faster response.

Table 9. Recognition rates (%) on EMODB reported in literature

| Reference | WAR | UAR | Reference | WAR | UAR | Reference | WAR | UAR |
|---|---|---|---|---|---|---|---|---|
| Haider *et al.,* 2020 [30] | N/A | 76.90 | Bhargava 2013 [78] | 80.60 | N/A | Özseven 2019 [96] | 84.62 | N/A |
| Zhao *et al.,* 2019 [61] | N/A | 79.70 | Badshah *et al.,* 2019 [79] | 80.79 | N/A | Deb 2017 [97] | 85.10 | N/A |
| Chen *et al.,* 2018 [62] | N/A | 82.82 | Zhang et al., 2013 [80] | 80.85 | N/A | Meng *et al.,* 2019 [98] | 85.32 | N/A |
| Song *et al.,* 2020 [24] | N/A | 86.19 | Wu *et al.,* 2011 [81] | 81.30 | N/A | Vasuki 2020 [99] | 85.54 | N/A |
| Sidorov *et al.,* 2016 [63] | 72.00 | N/A | Sun 2015 [82] | 81.50 | N/A | Sajjad 2020 [12] | 85.57 | N/A |
| Aghajani *et al.,* 2020 [64] | 72.10 | N/A | Sun *et al.,* 2015 [83] | 81.74 | N/A | Chen *et al.,* 2020 [23] | 85.61 | N/A |
| Li *et al.,* 2021 [22] | 72.19 | N/A | Xu *et al.,* 2015 [84] | 81.80 | N/A | Eyben *et al.,* 2015 [100] | 86.00 | N/A |
| Yüncü *et al.,* 2014 [65] | 72.30 | N/A | Vieira *et al.,* 2020 [28] | 81.80 | N/A | Issa *et al.,* 2020 [7] | 86.10 | N/A |
| Khan 2017 [66] | 72.34 | N/A | Man-Wai 2016 [85] | 81.86 | N/A | Singh *et al.,* 2020 [25] | 86.36 | N/A |
| Sinith *et al.,* 2015 [67] | 73.75 | N/A | Stuhlsatz *et al.,* 2011 [86] | 81.90 | N/A | Jiang *et al.,* 2019 [101] | 86.44 | 84.53 |
| Deb 2017 [68] | 73.90 | N/A | Wen *et al.,* 2017 [87] | 82.32 | N/A | Bakhshi *et al.,* 2020 [20] | 87.29 | N/A |
| Deb 2016 [69] | 74.40 | N/A | Lotfidereshgi 2017 [88] | 82.35 | N/A | Zhang *et al.,* 2017 [102] | 87.31 | 86.30 |
| Tao *et al.,* 2016 [70] | 74.46 | N/A | Sun 2017 [89] | 82.40 | N/A | Sun *et al.,* 2019 [103] | 87.55 | N/A |
| Kadiri *et al.,* 2015 [71] | 75.22 | N/A | Tzinis *et al.,* 2018 [90] | 82.40 | N/A | Hook *et al.,* 2019 [104] | 88.60 | N/A |
| Shirani *et al.,* 2016 [72] | 76.12 | N/A | Kalinli 2016 [91] | 82.70 | N/A | Jalili *et al.,* 2018 [105] | 88.80 | N/A |
| Bashirpour 2016 [73] | 76.60 | N/A | Hou *et al.,* 2020 [27] | 82.80 | 83.30 | Nguyen *et al.,* 2020 [106] | 89.00 | N/A |
| Sugan *et al.,* 2020 [21] | 77.08 | N/A | Daneshfar 2020 [5] | 82.82 | N/A | Sun 2019 [107] | 89.70 | N/A |
| Luengo 2010 [74] | 78.30 | N/A | Zhang 2021 [26] | 83.70 | N/A | Tuncer *et al.* 2021 [18] | 90.09 | N/A |
| Rintala 2020 [75] | 78.50 | N/A | Yi 2019 [92] | 83.74 | N/A | Er 2020 [10] | 90.21 | N/A |
| Wang *et al.,* 2020 [17] | 79.20 | N/A | Deb 2018 [93] | 83.80 | N/A | Sun 2020 [11] | 90.30 | N/A |
| Hassan 2012 [76] | 79.50 | N/A | Yi 2020 [15] | 84.49 | 83.31 | Farooq *et al.* 2020 [9] | 90.50 | N/A |
| Daneshfar *et al.,* 2020 [6] | 79.94 | 76.81 | Tawari 2010 [94] | 84.50 | N/A | **Proposed** | **91.29** | **88.95** |
| Zao 2014 [77] | 80.10 | N/A | Özseven 2018 [95] | 84.50 | N/A | | | |

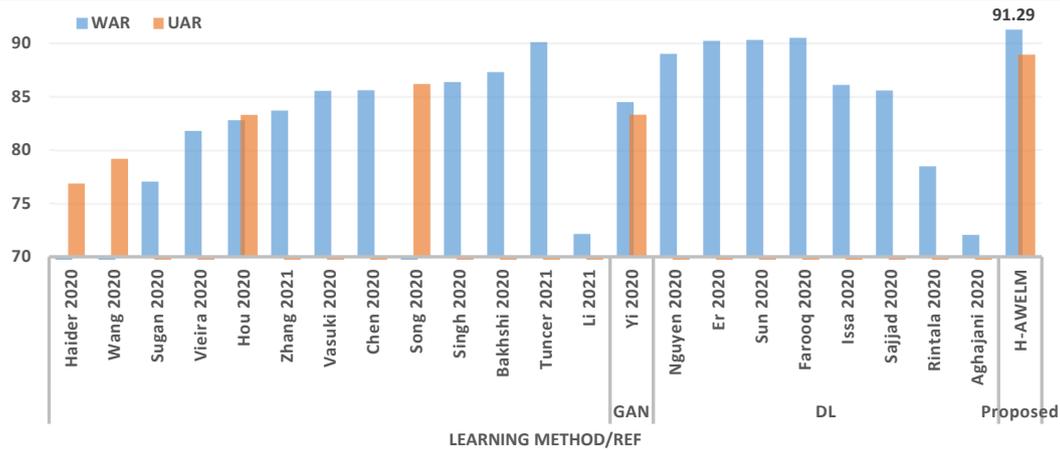

Fig. 14. Comparison of H-AWELM recognition rate with state of the art (EMODB)

Table 10. Recognition rates (%) on SAVEE reported in literature

| Reference | WAR | UAR |
|---|---|---|
| Haider *et al.,* 2020 [30] | N/A | 42.4 |
| Papakostas *et al.,* 2017 [108] | 44.00 | N/A |
| Liu *et al.,* 2018 [109] | 44.18 | N/A |
| Noroozi *et al.,* 2017 [110] | 45.51 | N/A |
| Vásquez-Correa *et al.,* 2016 [111] | 47.30 | N/A |
| Sun *et al.,* 2015 [83] | 50.00 | N/A |
| Sun 2017 [89] | 51.46 | 49.33 |
| Wen *et al.,* 2017 [87] | 53.60 | N/A |
| Tzinis *et al.,* 2018 [90] | 54.00 | 53.80 |
| Sugan *et al.,* 2020 [21] | 55.83 | N/A |
| Sinith *et al.,* 2015 [67] | 57.50 | N/A |
| Sun 2015 [82] | 58.76 | N/A |
| Daneshfar *et al.,* 2020 [6] | 59.38 | 55.00 |
| Zhang 2021 [26] | 60.16 | N/A |
| Daneshfar 2020 [5] | 60.79 | N/A |
| Nguyen *et al.,* 2020 [106] | 62.00 | N/A |
| Jiang *et al.,* 2019 [101] | 62.49 | 59.40 |
| Wang *et al.,* 2020 [17] | 66.20 | 81.8 |
| Farooq *et al.* 2020 [9] | 66.90 | N/A |
| **Proposed** | **66.94±0.65** | **67.90±0.62** |

Table 11. Recognition rates (%) on the IEMOCAP reported in literature

| Reference | WAR | UAR | Reference | WAR | UAR | Reference | WAR | UAR |
|---|---|---|---|---|---|---|---|---|
| Latif *et al.*, 2019 [112] | N/A | 60.23 | Hou *et al.*, 2020 [27] | 62.80 | 63.80 | Zhao *et al.*, 2019 [61] | 65.20 | 68.00 |
| Zong *et al.*, 2018 [113] | N/A | 64.80 | Tzinis *et al.*, 2018 [90] | 63.00 | 65.20 | Daneshfar 2020 [5] | 65.71 | 65.73 |
| Latif *et al.*, 2020 [16] | N/A | 66.70 | Li *et al.*, 2015 [121] | 63.20 | N/A | Deb 2018 [93] | 66.80 | N/A |
| Kwon 2020 [8] | N/A | 73.01 | Vieira *et al.*, 2020 [28] | 63.2 | N/A | Yi 2019 [92] | 66.80 | 62.83 |
| Ghosh *et al.*, 2016 [114] | 52.82 | N/A | Jiang *et al.*, 2019 [101] | 64.00 | N/A | Liu *et al.*, 2018 [109] | 67.10 | 66.20 |
| Xie *et al.*, 2019 [115] | 54.00 | N/A | Tzinis 2017 [122] | 64.16 | N/A | **Proposed** | **67.68** | **63.13** |
| Li *et al.*, 2020 [116] | 58.62 | 59.91 | Deb 2017 [97] | 64.20 | N/A | Yeh *et al.*, 2020 [126] | 69.00 | 70.10 |
| Zhao *et al.*, 2018 [117] | 59.70 | 60.10 | Chen *et al.*, 2018 [62] | 64.20 | N/A | Yi 2020 [15] | 71.45 | 64.22 |
| Huang *et al.*, 2018 [118] | 60.40 | N/A | Han *et al.*, 2018 [123] | 64.20 | 65.70 | Sun 2020 [11] | 71.50 | N/A |
| Li *et al.*, 2021 [22] | 60.83 | N/A | Issa *et al.* 2020 [7] | 64.30 | N/A | Li *et al.*, 2018 [127] | 71.75 | N/A |
| Xia 2015 [119] | 60.90 | 62.40 | Etienne *et al.*, 2018 [124] | 64.50 | 61.70 | Fan *et al.*, 2020 [128] | 73.02 | 65.86 |
| Zhao *et al.*, 2019 [61] | 61.90 | N/A | Fayek *et al.*, 2017 [125] | 64.80 | 60.90 | Daneshfar *et al.*, 2020 [6] | 74.80 | N/A |
| Mao *et al.*, 2019 [120] | 62.28 | 58.02 | Shirani *et al.*, 2016 [72] | 65.20 | N/A | | | |

Fig. 16 to 18 show the confusion matrices related to the EMODB, SAVEE, and IEMOCAP databases. According to the EMODB confusion table, *fear* emotion with 95.27% accuracy has the highest recognition rate compared to the other emotions. In addition, *bore* is most similar to *sad* compared to the other emotions due to the similarity of their prosodic features and glottal waveforms (91.89% to 5.43%). *Anger* is also the most similar to *happiness* among other emotions (93.91% to 6.08%), this similarity is also due to the similarity of their prosodic features (both of them have more energy than the other emotions) and consequently the similarity in their glottal waveforms too. In addition, in the SAVEE confusion matrix, the most similar emotion to *anger* is *happiness* (60.61% to 9.52%), this can also be seen in the confusion matrix for the IEMOCAP dataset too (58.64% to 23.66%). In fact, in the method presented in this paper, due to the use of features based on Mel frequency and perceptual linear coefficients inspired by human auditory structure, different emotions of *happiness*, *sad*, *anger*, *fear*, *neutral*, *bore* and *disgust* are well discernible and distinct in all three emotional datasets.

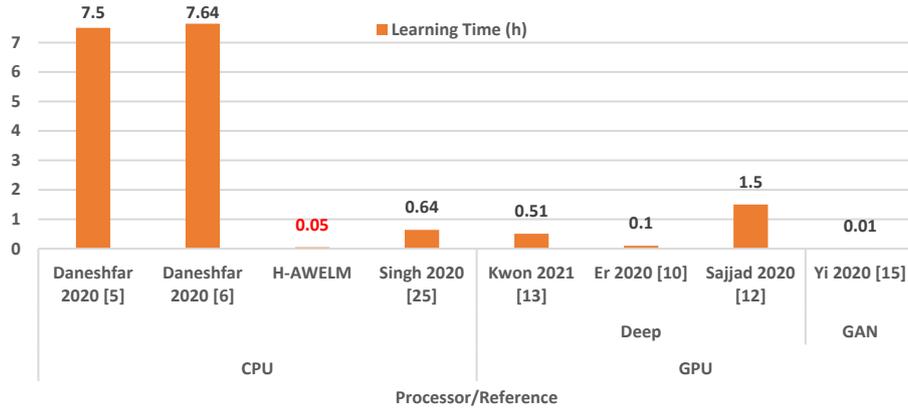

(a)

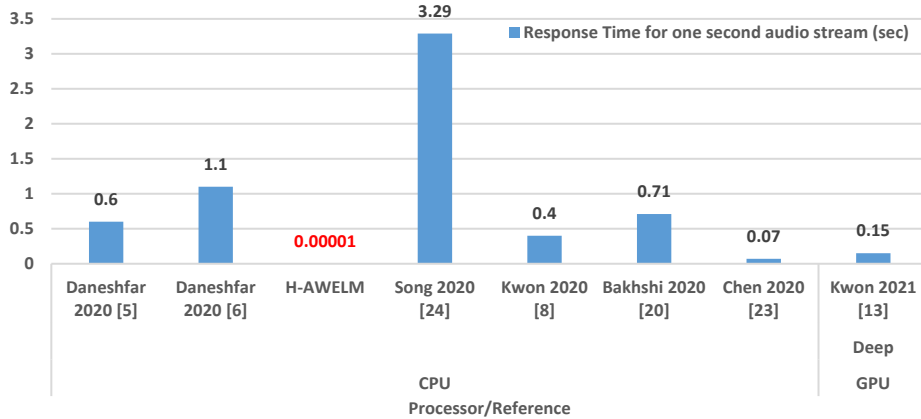

(b)

Fig. 15. Comparison (a) Execution time and (b) Response time with state of the art (EMODB)

|       | F     | D    | H    | B    | N    | S    | A    |
|-------|-------|------|------|------|------|------|------|
| Fear  | 95.27 | 0    | 3.72 | 0    | 0    | 0    | 1    |
| Disgust | 0   | 84.8 | 0    | 11   | 0.1  | 0    | 4.1  |
| Happiness | 3.03 | 0 | 85.86 | 0   | 1.25 | 0    | 9.83 |
| Bore  | 0     | 2    | 0    | 91.89 | 0.66 | 5.43 | 0   |
| Neutral | 0.7 | 0    | 1.43 | 2.08 | 94.97 | 0.80 | 0  |
| Sad   | 0.90  | 0.90 | 0    | 6.51 | 0    | 91.66 | 0  |
| Anger | 0     | 0    | 6.08 | 0    | 0    | 0    | 93.91 |

Fig. 16 Confusion matrix of the proposed method for EMODB

|    | A    | D    | F    | H    | N    | SA   | SU   |
|----|------|------|------|------|------|------|------|
| A  | 60.61 | 9.14 | 8.33 | 9.52 | 2.97 | 4.06 | 5.32 |
| D  | 3.9  | 62.01 | 6.10 | 5.21 | 10.58 | 5.36 | 6.71 |
| F  | 2.18 | 2.32 | 71.50 | 7.29 | 0.31 | 5.01 | 11.35 |
| H  | 8.60 | 2.29 | 2.81 | 66.67 | 1.33 | 4.99 | 13.28 |
| N  | 1.64 | 4.52 | 1.45 | 1.25 | 79.29 | 11.51 | 0.30 |
| SA | 0    | 5.02 | 10.48 | 3.95 | 8.70 | 67.21 | 4.62 |
| SU | 0    | 1.15 | 19.44 | 10.81 | 0    | 0.63 | 67.95 |

Fig. 17 Confusion matrix of the proposed method for SAVEE

|          | Anger | Happiness | Neutral | Sad   |
|----------|-------|-----------|---------|-------|
| Anger    | 58.64 | 23.66     | 14.50   | 3.18  |
| Happiness| 9.12  | 65.23     | 13.94   | 11.69 |
| Neutral  | 7.47  | 16.14     | 68.83   | 7.54  |
| Sad      | 13.79 | 22.51     | 5.49    | 58.18 |

Fig. 18 Confusion matrix of proposed method for the IEMOCAP

Compared to the other proposed architectures, [17] has achieved 79.12% accuracy using features based on wavelet packet analysis, along with the linear SVM classifier on the EMODB. Also, an accuracy of 76.9% was obtained in [30] using a feature selection method based on Fisher criteria on the same dataset. Paper [106] achieved 89% and 62% recognition rates on the EMODB and SAVEE datasets respectively by reducing inter-class variance and increasing intra-class variance, in order to train SER systems on emotional datasets with incomplete data and knowledge transfer between them. One reason for the high accuracy of H-AWELM compared to the mentioned studies is the use of a multi-layer architecture based on an ELM-AE to discover and extract more complex features from the input features. In addition, using the glottal waveform and SFCC, ICMC, CQCC features increase the ability to distinguish between different emotions and even recognize the same emotions for both groups of different speakers and the same speakers (see Table 9 to 11). While [17] only using features based on wavelet packet analysis, it has not been able to distinguish seven different emotions precisely (low recognition rate of *fear* 55.05% and *neutral* 56.96%, compared to 95.27% and 94.97% in H-AWELM). In addition, the paper [25] has failed to detect the emotion of *sad* (8.86%) in the EMODB (although it has been successful in recognizing the SAVEE dataset emotions). As well as the paper [24] has produced a SER system with a small recognition rate for *anger* (83.26%) compared to (93.91%) in H-AWELM, by extracting a high-dimensional feature vector (6373) and using a proposed feature selection algorithm to reduce it.

Also, the recognition rate of the proposed H-AWELM network on the SAVEE and IEMOCAP datasets are comparable to many of the recent state of the works so far in the same experimental conditions (Table 10 to 11). However, recent results published by the authors [5], using the pQPSO dimension reduction method and the Gaussian classifier, have achieved higher accuracy on this emotional database. It seems that the proposed weighted classifier H-AWELM in this case, has not been able to more accurately detect different emotions on IEMOCAP. However, in the proposed architecture, due to the use of ELM, the response time of the algorithm is less than many other classifiers, including the Gaussian classifier.

In the following, Table 12 shows the recognition rate of the H-AWELM network for ten different folds of EMODB. According to the features used in this study, the emotions expressed by each speaker (with any gender), are well recognizable. In fact, the emotions expressed by the female speaker (92.55%) and the emotions expressed by the male speaker (92.04%) are recognized well by the proposed architecture too. Actually, using the ICMC, CQCC and SFCC features that have a good ability to confirm and recognize speakers, is one of the reasons for recognizing emotions in different speakers. In addition, the use of glottal waveform features has led to better recognition and differentiation of different genders' emotions too.

Despite the high accuracy of the proposed architecture, one of its problems and limitations is the high memory consumption due to the hierarchical structure and high dimensions of the ELM's output layer weights due to the high dimensions of the input feature vector. In the H-AWELM model, the memory consumed during the training, is equal to the sum of the weights of the hierarchical ELM-AE's layers ($\beta_i$) with the output layer weights of the ELM classifier. Fig. 19 shows the memory consumption in the various H-AWELM network configurations. The best result obtained on the EMODB dataset, requires 3.232 GB of memory consumption. Due to the memory consumed during training, it will not be possible to further examine the results by different configurations of this network.

Table 12. Recognition rates (%) on the EMODB for different speakers and genders

| fold no (gender) | WAR |
|---|---|
| 1 (m) | 94.55 |
| 2 (f) | 88.82 |
| 3 (f) | 89.97 |
| 4 (m) | 90.25 |
| 5 (m) | 93.68 |
| 6 (m) | 89.93 |
| 7 (f) | 90.58 |
| 8 (f) | 92.54 |
| 9 (m) | 91.79 |
| 10 (f) | 90.85 |
| Mean (m) | 92.04 |
| Mean (f) | 90.55 |

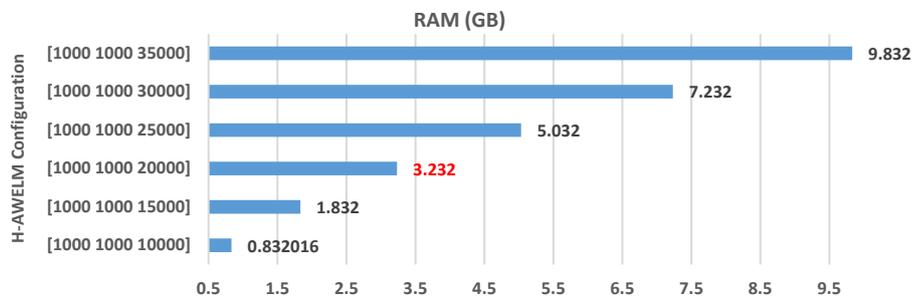

Fig. 19 Comparison of memory consumption in different H-AWELM configurations

### 5-3 Multilingual Results

Recently, interest in SER systems has been increased from monolingual scenarios to multilingual ones so these systems are able to recognize speech emotions across languages like any humans. On the one hand, this interest can promote full development of SER systems in the real world and provide an opportunity to examine similarities between languages so that other speech instances from different datasets can be used in the absence of training data in emotional models. This is also useful when emotional dataset samples are low; therefore, the other datasets will be used in the training set to boost efficiency and recognition rates.

In order to investigate the performance of the proposed system in a multilingual mode, exploiting datasets with different languages can be very useful and effective. It should be noted that EMODB is in German and SAVEE and the IEMOCAP are in British English and American English; respectively. To evaluate the proposed system in a cross-speaker mode, all datasets must be included in the training set and only one speaker will be in the test set. The accuracy rate for each dataset will be also equal to the average of the results for all speakers of that particular dataset. Since all three datasets will be employed as training and test sets in this experiment, they should have the same level of participation and emotions. For this purpose, the following four emotions of *neutral*, *happiness*, *anger*, and *sad* presented in all corpora will be used. Table 13 shows details of the utterances' numbers chosen from each corpus.

The proposed system is only useful if it can produce comparable results in a multilingual mode compared with a monolingual system. As cited in Table 14, the recognition rate of 75.94% is achieved in the multilingual mode for EMODB corpus compared with 94.16% in monolingual case. This recognition rate for SAVEE corpus in multilingual mode is 47.23% compared with monolingual accuracy of 75.33%. But the difference for the IEMOCAP dataset is smaller than the other two, 66.11% vs. 67.68% in multilingual and monolingual cases; respectively. The reason for this small difference is that the IEMOCAP corpus has more utterances than the other two datasets; therefore, it has a larger volume of training set and the difference of accuracy rates in multilingual and monolingual modes is less than that of two other corpora.

Table 13. Four emotional corpora and their emotions and numbers of utterances

| CORPUS | LANGUAGE | EMOTION | | | | TOTAL |
|---|---|---|---|---|---|---|
| | | Neutral | Happiness | Anger | Sad | |
| EMODB | German | 60 | 60 | 60 | 60 | **240** |
| SAVEE | English | 60 | 60 | 60 | 60 | **240** |
| IEMOCAP | English | 100 | 100 | 100 | 100 | **400** |

Also, it was expected that the accuracy rate in the multilingual mode would be better than the monolingual one, because the data in the training set has increased. However, according to the results presented in Table 14, it seems useless to add English data to the German EMODB training set or adding German data to the English SAVEE and IEMOCAP training sets. The reason for this decrease in accuracy rate is that the extracted features are dependent on the environment and specified languages; however, it is better to use features that are independent of environmental and linguistic conditions in future work.

Table 14. Multilingual vs. monolingual results (%)

| | Train | Test | WAR |
|---|---|---|---|
| **Monolingual Train & Test** | EMODB | EMODB | 94.16+0.43 |
| | SAVEE | SAVEE | 75.33±1.09 |
| | IEMOCAP | IEMOCAP | 67.68±0.33 |
| **Multilingual Train & Monolingual Test** | EMODB, SAVEE, IEMOCAP | EMODB | 75.94±1.72 |
| | | SAVEE | 47.23±1.19 |
| | | IEMOCAP | 66.11±0.31 |

## 5-4 Statistical Analysis of H-AWELM

In this section, the proposed method is evaluated in terms of speed, accuracy, generality and output richness criteria. Statistical analysis of the presented results is used to organize, summarize and further compare it, by descriptive statistical methods.

### 5-4-1 T-Test

The purpose of this section is to compare the best results reported in the paper with the best results reported so far in other references through t-test evaluation. Table 15 shows the best recognition rate ever obtained for each emotional data set in LOSO mode.

Table 15. Comparison of p-Values for the best recognition rate ever achieved in LOSO mode

| Dataset | EMODB | SAVEE | IEMOCAP |
|---|---|---|---|
| H-AWELM WAR | 91.29±0.43% | 66.94±0.65% | 66.94±0.65% |
| The best competitor WAR | Farooq *et al.*, 2020 [9] 90.50% | Farooq *et al.*, 2020 [9] 66.90% | Daneshfar 2020 [5] 74.80±0.11% |
| p-value | $p < 0.01$ | $p > 0.01$ | $\boldsymbol{p > 0.01}$ |
| Hypothesis confirmed | $\mu > \mu_0$ | $\mu \leq \mu_0$ | $\boldsymbol{\mu \leq \mu_0}$ |

Also, the results of p-value evaluation using t-test with confidence value ($\propto$) of 0.01 are also shown in Table 15. In this test, we want to check whether the mean of the best recognition rate obtained by the H-AWELM ($\mu$) on emotional datasets is less than or equal to the mean of the best recognition rate obtained by the competitor ($\mu_0$) (hypothesis zero or $H_0$), or not (opposite hypothesis or $H_a$)? This is shown in relation (52),

$H_0: \mu \leq \mu_0$

$H_a: \mu > \mu_0$ (52)

Considering the p-values for EMODB, it can be concluded that the recognition rate obtained by the H-AWELM, with a high confidence value, has a significant difference compared to the best method presented so far on this dataset and in LOSO mode. In the case of SAVEE dataset, although the mean obtained in this study is higher than the average of the competitor, this superiority cannot be claimed with high reliability (*p = 0.46*). Also, in the case of the IEMOCAP method, the average obtained from the proposed method was lower than those published earlier by the authors.

### 5-4-2 Dataset's Recognition Rates

As shown in Fig 20 and in comparison, with those published earlier by the authors, the highest mean and median value among the three different emotional datasets is related to EMODB and the lowest value is related to SAVEE. Due to the small number of samples of the training set compared to the samples of the test set in SAVEE dataset compared to EMODB and IEMOCAP, a significant decrease in the recognition rate in LOSO mode is evident in this dataset. The H-AWELM method also had the highest recognition rates on EMODB and SAVEE, while [5] method produced better results on IEMOCAP.

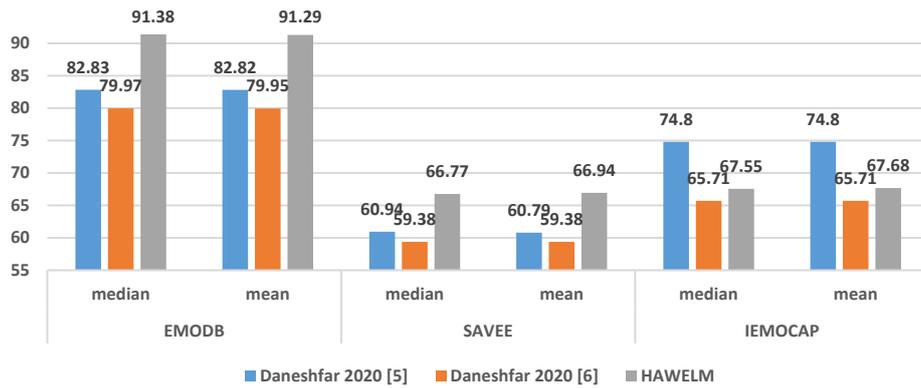

Fig. 20 Comparison of the datasets' recognition rate of the presented methods by the authors using measures of central tendency

One of the reasons for the high recognition rate of IEMOCAP using the pQPSO method in [5], is the high utterances' number (5531 samples) in this dataset compared to EMODB and SAVEE datasets (535 and 480 samples respectively) which has achieved better results by using the GMM statistical classifier with more training samples. In addition, these results indicate that the long-term features used in the H-AWELM method have improved results on the acted datasets (EMODB and SAVEE). However, the short-term features in [5] have yielded better results on IEMOCAP with both improvisations scenarios. In addition, the short-term features used in [5] are suitable for longer speech formats (the utterances' average length of IEMOCAP in these experiments is $4.73 \pm 0.18$ seconds). While the long-term features in the H-AWELM are more suitable for speech formats with shorter average length (the utterances' average length of EMODB and SAVEE is 2 to 3 and $3.85 \pm 0.33$ seconds respectively). Therefore, it can be concluded that the presented method in [5] will have a higher recognition rate on datasets with more samples but will distinguish less emotions (4 different emotions in IEMOCAP in comparison to 7 emotions in EMODB and SAVEE). While the long-term features and the H-AWELM method has better recognition rate on the shorter-length datasets and will differentiate more emotions (7 emotions in EMODB and SAVEE in comparison to 4 different emotions in IEMOCAP).

### 5-4-3 Emotions' Recognition Rates

Fig. 21 shows the recognition rate of different emotions for the proposed methods by the authors, and for different emotion sets. According to Fig. 21 (a), all the proposed methods identified *anger* and *sad* emotions at a higher rate on the EMODB dataset, due to the large number of its samples. This is quite evident, for the *neutral* emotion, which has many samples in SAVEE and IEMOCAP datasets too (Fig. 21 (b) and Fig. 21 (c)). In addition, *sad* and *anger* emotions with fewer samples in IEMOCAP has lower recognition rates (Fig. 21 (c)).

According to Fig. 21 (d), *bore*, *neutral*, *happiness*, and *fear* emotions has higher recognition rates using the long-term features and the H-AWELM method. While *neutral*, *anger*, *disgust*, and *surprise* emotions have better results with short-term features and the pQPSO method [5]. Emotions with instantaneous changes (such as *anger* and *surprise*) appear to have higher recognition rates using short-term features and the GMM statistical classifier [5]. Also, emotions that change slightly over time (such as *bore* and *neutral*) are more recognizable with long-term features and H-AWELM method. In addition, because the *disgust* emotion has fewer acoustic features and low sample numbers in emotional dataset, its detection seems to be more complex than other emotions, and the most researches did not report its recognition rate.

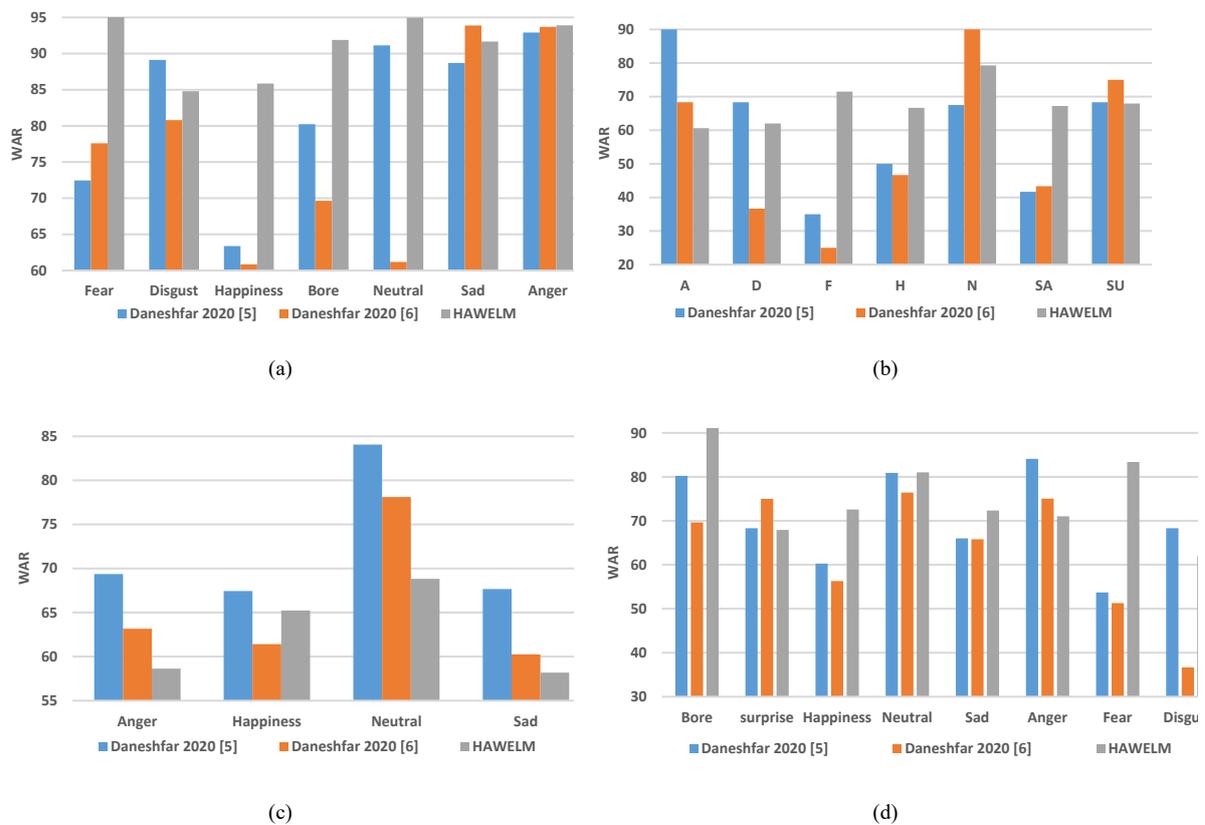

Fig. 21 Comparison of the emotions' recognition rates of the presented method by the authors

### 5-4-4 Correlated Emotions' Recognition Rates

Fig. 22 provides a comparison on the effect of H-AWELM features, on distinguishing the correlated emotions for the EMODB dataset (these emotions have already been identified in Fig. 1). In this evaluation, an experiment was performed with H-AWELM model without SFCC, ICMC, CQCC, GBFB, and glottal features, respectively, and the recognition rate of the correlated emotions in each case, are shown in Fig. 22 (a to c). As it is clear in Fig. 22 (a), the absence of any of these features has a negative impact on the distinction of correlated pairs (*bore*, *sad*). The SFCC and glottal waveform features have the most effects on differentiation of these two emotions than the other features. Also, GBFB and CQCC features has a little effect on distinguishing of (*fear*, *disgust*) pair (Fig. 22 (b)), while the glottal waveform features have the greatest effect on (*fear*, *disgust*) and (*anger*, *happiness*) recognition rates too.

Fig. 23 shows a comparison between the correlated emotions' classification error, compared to the most recently presented results for the EMODB dataset.

For each correlated pair ($C_1$, $C_2$) the classification error of ($C_1/C_2$) will be provided by,

$$Classification\ Error\ (C_1/C_2) = \frac{\#false\ negative\ of\ C_1\ (misclassifying\ C_1\ as\ belonging\ to\ C_2)}{\#\ true\ positive\ of\ C_2} \qquad (53)$$

Fig 23. (a) shows the classification error of the recently methods for (*bore, sad*) emotions; Fig. 23 (b) shows the classification error of (*fear, disgust*) emotions; Fig. 23 (c) shows classification error of (*anger, happiness*) emotions and Fig. 23 (d) shows the total number of correlated emotion classification errors.

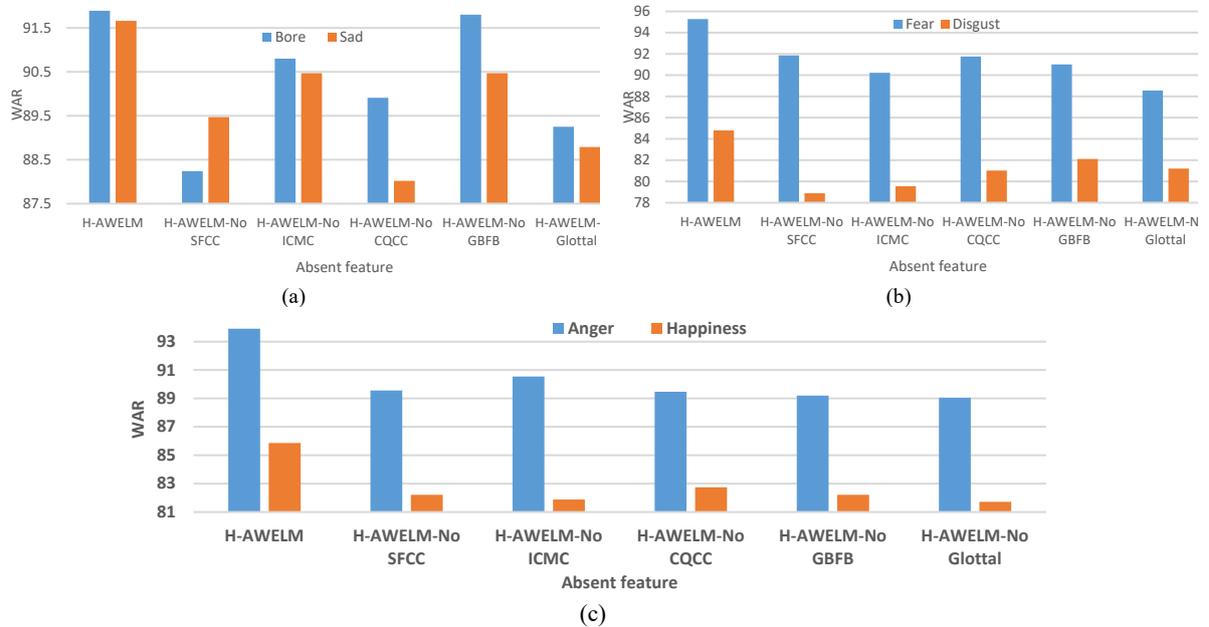

Fig. 22. Comparison of the effect of different features on the recognition rate (WAR) of EMODB correlated emotions (a) (*bore, sad*) (b) (*fear, disgust*) (c) (*anger, happiness*)

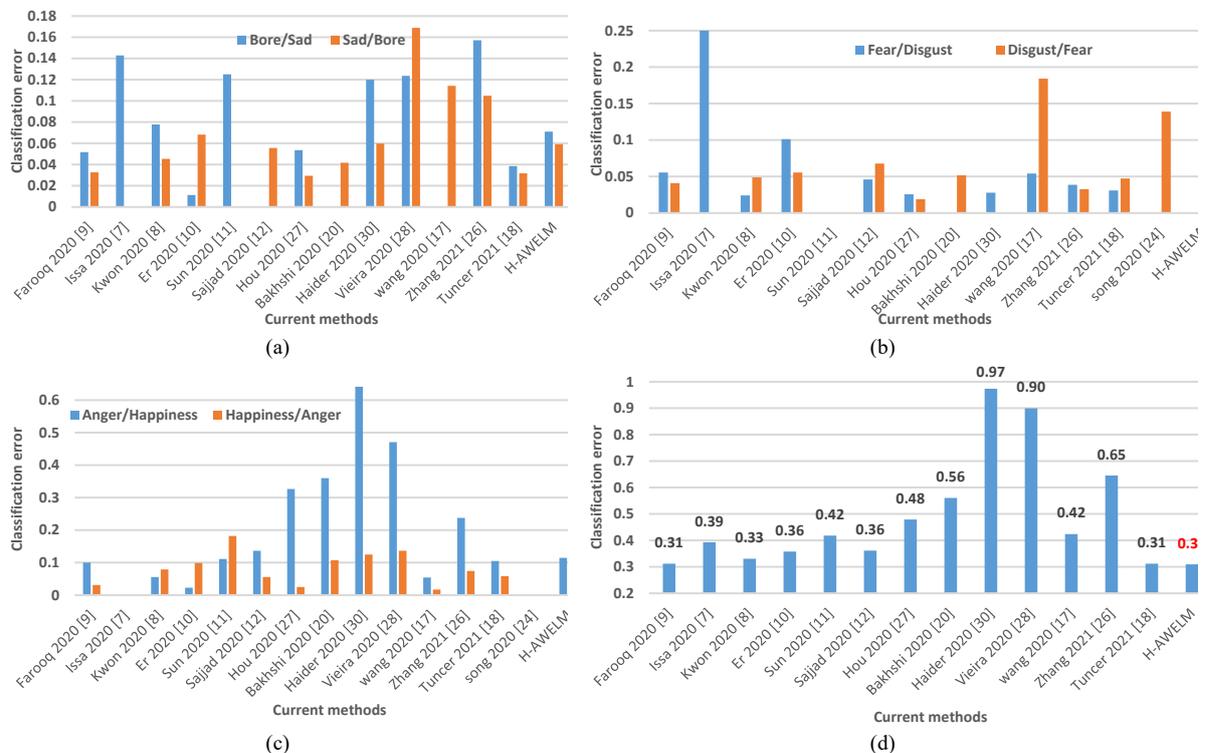

Fig. 23. Comparison of EMODB correlated emotion classification error in the proposed method and recent researchs (a) *bore* and *sad* (b) *fear* and *disgust* (c) *anger* and *happiness* and (d) total errors

According to Fig. 23 (d), many recent methods have failed to distinguish all of the correlated emotions well at high rates, and a classification error has been occurred. Compared to the recent works, the H-AWELM method has produced a high recognition rate for similar and correlated emotions. However, the method proposed by Tuncer [18] has also been able to differentiate correlated and similar emotions at a higher rate than other recently methods.

## 6- Discussion

In this paper, a new method for SER has been proposed based on the H-ELM with a new weighting method as a classifier that makes the final system more accurate in contrast to the recently weighting algorithms as well as many recently published SER systems. Using the glottal waveform features -which will have a great impact on the speaker's speech style, behavior and emotional states- the ability to identify and recognize similar emotions has been done well. In addition, the use of CQCC, ICMC and SFCC features, which have not been used in SER systems so far, allows better recognition and differentiation of the emotions of different speakers with different genders. This paper also, utilizes the spectro-temporal GBFB and SGBFB features of input speech signal which are efficient in reducing redundancy between feature components and have good discriminative properties in both spectral and temporal domains.

In this research, various experiments have been conducted to find appropriate features, statistics, feature selection numbers, and classification methods with their kernels and structures. For example, the results of the experiments indicate that the set of features has the most impact on the accuracy rate of the system including MFCC, PLPC, CQCC, ICMC, and SFCC along with their first- and second-order derivatives and also GBFB and SGBFB features. Likewise, a good set of long-term statistics (i.e. mean, SD, skewness, and kurtosis), suitable for extracting features related to utterances, was obtained through numerous experiments. Due to the high dimensional feature vector in the proposed algorithm, classical feature selection methods and a new quantum-inspired dimension reduction method with a variety of feature selection numbers could be used in which, according to the experiments, the best selected feature numbers with the best accuracy rates were 3000 and 5000. In addition, the effects of various classifiers including different SVM structures, deep NNs, and different ELM-based methods with various kernels and structures have been additionally investigated. Finally, the H-ELM method has been selected with hyperbolic tangent activation function as kernel and a significant simple structure which learns considerably faster than existing learning methods.

With respect to the new weighting method, the proposed hierarchical adaptive weighted ELM network is able to minimize the total misclassification costs with low computational complexity and yield a more accurate model. To prove this, various experiments have been done on four different recent weighting methods compare to the proposed weighting algorithm on three different emotional speech databases in both 3000 and 5000 number of selected features.

Despite the above-mentioned advantages and as it is clear from the experimental results, the ELM shows a very sensitive performance regarding to the number of hidden layers and the number of neurons in each layer, so it is necessary to use a faster method to find the optimal number of layers and neurons. In addition to the optimal neuron numbers, due to the proposed hierarchical structure, memory is also a critical issue for this strategy and high memory requirement is a major drawback of the proposed method that can increase the implementation costs. Then, in this case, the training phase can be executed on parallel computers to fix high memory usage issue. Another limitation presented in this study is related to exploiting the environmental and language-dependent features that are not fully robust and make the results in multilingual mode not as good as the monolingual one. To avoid this, it is better to improve feature sets to be language-independent that are more robust features in the future.

All the mentioned experimental results on three famous emotional databases have been presented to verify the efficacy of the proposed method and to show improved and better results compared with the recently-published works.

For future work, the authors would like to establish deep NNs instead of layers of the proposed H-AWELM, to improve and use more robust and language-independent features, and to examine more emotional corpora in both multilingual and cross-corpus cases. The purpose of this experiment compared with previous ones is to find the relationship between emotions and languages as well as speech features among languages.

# Conclusion

In this paper, a three-stage system was proposed for speech emotion recognition. In the first stage, long-term statistics are extracted from prosodic, spectral and spectro-temporal features, both from the speech signal and glottal waveform signal, and a feature vector with a very large number of dimensions is formed. In the second stage, the dimensions of this feature vector are reduced, using classical feature selection methods as well as a proposed new quantum-inspired method. In the third stage, a deep ELM-based classifier is used that has three steps: sparse random unsupervised feature learning, orthogonal random feature mapping, and generalized Tikhonov regularization-based supervised learning.

The results obtained in this paper are as follows:

- In feature extraction phase, prosodic feature vectors have the lowest accuracy and spectro-temporal feature vectors have the highest accuracy in emotion classification.
- In long-term statistics extraction phase, classical first- to fourth-order statistics (mean, standard deviation, skewness and kurtosis) were more efficient than harmonic and geometric statistics, percentile and zero crossing rate ones.
- In the classical feature selection phase, the mRMR method was more accurate than the other methods.
- In the dimension reduction phase, a proposed new method inspired by quantum theory was used.
- In the emotion classification phase, the three-stage deep ELM-based classifier was more accurate than other types of ELM, as well as SVM classifiers and deep neural network classifiers.
- In the emotion classification phase, to deal with the problem of class imbalance in emotional databases, the proposed new weighting method was more accurate than other weighting methods.
- According to our latest information, the proposed system outperforms other state-of-the-art systems on two well-known emotional databases.

Currently, the authors of the paper are using time-recurrent classifiers based on reservoir computing (such as echo state networks (ESN)) to better model the sequence of emotional acoustic events over time. The following can be suggested as future works: use of features such as predictive features [129], system extension for telephone and narrowband channels using bandwidth expansion technique [130] (the current system is designed for broadband channels), use of new simulated annealing-based algorithms such as MiGSA [131] for dimension reduction and finally using state-switching acoustic models such as time-inhomogeneous hidden Bernoulli models (TI-HBM) [132] for emotion classification.